\documentclass{iopjournal}
\usepackage{amsmath}
\usepackage{amssymb}
\usepackage{ragged2e}
\usepackage[
  left=2.2cm,
  right=2.2cm,
  top=4.5cm,
  bottom=2.5cm
]{geometry}

\begin{document}
\pagestyle{plain}

\title{Extreme Quantum Cognition Machines for Deliberative Decision Making}

\author{Francesco Romeo$^{1,2}$\orcid{0000-0001-6322-7374},
Jacopo Settino$^{3,4,*}$\orcid{0000-0002-7425-4594}}

\affil{$^{1}$Dipartimento di Fisica ``E. R. Caianiello'', Universit\`a degli Studi di Salerno, Via Giovanni Paolo II, I-84084 Fisciano (SA), Italy}

\affil{$^{2}$INFN, Gruppo Collegato di Salerno, Via Giovanni Paolo II, I-84084 Fisciano (SA), Italy}

\affil{$^{3}$Dipartimento di Fisica, Universit\`a della Calabria, I-87036 Rende (CS), Italy}

\affil{$^{4}$INFN, Gruppo Collegato di Cosenza, I-87036 Arcavacata di Rende (CS), Italy}

\affil{$^{*}$Author to whom any correspondence should be addressed.}

\email{jacopo.settino@unical.it}


\begin{abstract}
\justifying
We introduce Extreme Quantum Cognition Machines, a class of quantum learning architectures for deliberative decision making designed to operate under coarse-grained, ambiguous, and potentially contradictory information. Inspired by the quantum cognition paradigm, Extreme Quantum Cognition Machines are closely related to quantum extreme learning and quantum reservoir computing, where fixed quantum dynamics generates a nonlinear feature map and learning is confined to a linear readout. A dynamical attention mechanism, implemented through an input-dependent interaction term in the Hamiltonian, modulates the quantum evolution and biases the resulting feature embedding toward task-relevant correlations. The approach is validated on linguistic classification tasks, which serve as paradigmatic examples of deliberative inference. Hardware-compatible quantum implementations of the proposed framework are discussed, together with potential applications in symbolic inference, sequence analysis, anomaly detection, and automatic diagnosis, with direct relevance to domains such as biology, forensics, and cybersecurity.
\end{abstract}

\section{Introduction}
\justifying

The development of learning machines inspired by physical and cognitive principles has a long and interdisciplinary history, spanning statistical physics, neuroscience, and artificial intelligence \cite{rosenblatt1958perceptron,hopfield1982neural,amit1985spin,ackley1985learning,mcculloch1943logical}. Early attempts to formalize learning and cognition in physical terms can be traced back to the pioneering work of Eduardo R. Caianiello, who introduced fundamental ideas on neural networks and associative memory well before the modern resurgence of artificial intelligence \cite{caianiello1961outline,NobelPhysics2024Background}. These contributions anticipated several key concepts underlying contemporary neural architectures and were later acknowledged as part of the intellectual lineage leading to modern models of collective memory and learning, including those associated with John Hopfield \cite{little1975statistical,willshaw1969non,kohonen2009correlation,hopfield1984neurons,amit1985storing}.

Over the past decades, neural networks have evolved into highly expressive and scalable models, forming the backbone of modern machine learning. Their success, however, has come at the price of increasingly complex training procedures, large computational costs, and limited interpretability. These limitations have motivated the exploration of alternative paradigms in which learning is simplified, localized, or structurally constrained, while expressive power is retained through rich internal dynamics. From this perspective, models based on linear training layers coupled to fixed nonlinear transformations have attracted renewed attention, both for their computational efficiency and for their potential suitability in adaptive and agent-based artificial intelligence systems \cite{lecun2015deep,schmidhuber2015deep,rumelhart1986learning,schwartz2020green,henderson2020towards,rudin2019stop,maass2002real,jaeger2004harnessing,huang2006extreme,tanaka2019recent}.

Reservoir computing and extreme learning machines embody this philosophy. In these approaches, a high-dimensional dynamical system—the reservoir—acts as a nonlinear feature map, while learning is restricted to a linear readout layer \cite{Fujii2017,Schuld2021,Xiong2025,Settino2024,Salatino2026,Ghosh2021,Mujal2021,Finocchio2024,Gyurik2025,Ghosh2021a,Settino2026,Romeo2026}. This separation dramatically simplifies the training procedure and enables rapid adaptation, making such architectures attractive for tasks involving streaming data, nonstationary environments, or multiple interacting agents. In recent years, these ideas have been extended to the quantum domain, giving rise to quantum reservoir computing and quantum extreme learning schemes, where quantum dynamics provides the nonlinear embedding of classical or quantum inputs.

The motivation for this extension lies in the possibility of exploiting the rich dynamical correlations and high-dimensional Hilbert-space structure of quantum systems to generate expressive nonlinear feature representations, while retaining the computational simplicity of linear training. In this perspective, quantum dynamics provides an alternative physical resource for feature generation, without necessarily implying a quantum computational advantage.

In parallel, the paradigm of quantum cognition \cite{BusemeyerBruza2012,Pothos2013,Khrennikov2004,Aerts2009} has emerged as a powerful framework to model decision making, contextual reasoning, and cognitive processes that elude classical probabilistic descriptions \cite{Accardi1981,Gudder1993}. In quantum cognition, cognitive states are represented as quantum states, questions and categories as observables, and decision outcomes/tendencies as projective measurements or expectation values. This formalism naturally accounts for contextuality, order effects, and ambiguity, which are intrinsic features of human deliberation and of many real-world decision problems.

In this work, we bring these two lines of research together. Our aim is to integrate the conceptual insights of quantum cognition with the architectural advantages of quantum extreme learning, in order to construct learning systems explicitly designed for quantum deliberation. By quantum deliberation we mean a decision-making process, implemented by means of the quantum cognition paradigm, tolerant to contradictory, context-dependent inputs. This setting is particularly relevant for hard classification tasks with noisy labels, ambiguous training data, and internal inconsistencies, where traditional learning paradigms often struggle.

The specific motivation for introducing EQCMs is to provide an operational realization of quantum deliberation in which the quantum state represents the available information, fixed quantum dynamics generates context-dependent correlations, and a learned linear combination of observables defines the deliberative criterion. This construction preserves the simplicity of extreme learning while providing a direct interpretation of the decision-making process within the quantum cognition framework.

The following sections will develop this idea in detail, moving from the conceptual foundations to algorithmic implementation, benchmark validation, and considerations on hardware compatibility.

The work is organized as follows. In Sec.~\ref{sec:QC} we introduce the quantum cognition framework and formalize the notion of deliberative decision making in terms of density matrices, observables, and expectation values. In Sec.~\ref{sec:QEL} we define the EQCM architecture within a quantum extreme learning paradigm, specifying the role of fixed quantum dynamics and linear readout for hard deliberative tasks. Sec.~\ref{sec:ALG} provides the full algorithmic implementation, including the maximum-entropy encoding of classical inputs into quantum states, the structure of the Hamiltonian dynamics, and the ridge-regression training procedure. In Sec.~\ref{sec:BENCH} we assess the performance of the proposed architecture on two symbolic linguistic benchmarks designed to exemplify hard deliberation, and we analyze the impact of encoding strategies and dynamical attention. Sec.~\ref{sec:HF} reformulates the model in a hardware-compatible form based on local Ising-type interactions and nearest-neighbour measurements, demonstrating its suitability for NISQ devices.
A formal discussion about the spectral properties of the deliberative operator is reported in Sec. \ref{sec:DelOp}.\\
Finally, Sec.~\ref{sec:CONCL} discusses conceptual implications, scalability, and prospective applications. Technical details concerning
the maximum-entropy construction underlying the initialization
of the quantum state and the performance metrics derived
from the confusion matrix are presented in Appendices \ref{AppA}
and \ref{AppB}, respectively. Appendix \ref{AppC} investigates the performance scaling of the EQCM with the quantum feature vector dimension
and provides a quantitative comparison with classical Extreme Learning Machines.

\section{Quantum cognition and deliberative decision making}
\label{sec:QC}

Quantum cognition \cite{BusemeyerBruza2012} is a theoretical framework developed to model decision-making and cognitive phenomena that systematically violate the axioms of classical probability theory. Its motivation does not stem from the assumption that the brain operates as a quantum physical system, but rather from the recognition that the mathematical structure of quantum probability \cite{Accardi1981,Gudder1993} provides a more general and flexible language to describe contextual, sequential, and ambiguous reasoning processes. In particular, quantum cognition adopts the formalism of Hilbert spaces and operators as an extension of classical probability based on set membership and measure theory.

Historically, the quantum formalism emerged from the effort to understand radiation–matter interaction, leading physicists to a probabilistic theory grounded in subspaces of a Hilbert space rather than in events represented as subsets of a sample space. In this setting, probabilities are not assigned to mutually exclusive events, but to projections onto subspaces, and the order in which questions are asked becomes operationally relevant. Quantum cognition exploits this mathematical structure to model cognitive phenomena such as context dependence, order effects, and interference, which are difficult or impossible to capture within a classical probabilistic framework.

Within quantum cognition, the mental state of an agent is represented by a quantum state in a Hilbert space. Questions, concepts, or evaluative categories are associated with Hermitian operators acting on this space. A response to a given question, or a deliberative act, is modeled as a measurement of the corresponding operator. As in quantum mechanics, the measurement process updates the mental state, projecting it onto an eigenstate of the operator associated with the question. When successive questions correspond to non-commuting operators, the final outcome depends on the order in which the questions are posed, naturally accounting for experimentally observed order and context effects in human decision making.

A central aspect of this framework is the use of a non-commutative probability theory. In contrast to classical probability, where joint distributions are always well defined, quantum probability admits observables for which no joint probability distribution exists, implying that their outcomes cannot be meaningfully assigned simultaneously, even in principle. This feature provides a principled way to describe incompatible cognitive evaluations and competing interpretations of the same stimulus. Importantly, the expectation value of an operator on the mental state represents a deliberative tendency, rather than a binary outcome, and can be interpreted as the aggregate result of multiple potential responses.

In this work, we focus on a class of problems that we refer to as \emph{deliberative tasks}. A deliberative task arises when a decision maker is required to assign a binary label to a given case, such as guilty or innocent, healthy or diseased, anomalous or ordinary, based on the evaluation of a collection of $m$ decision elements. Each element takes values in a finite and discrete abstract alphabet, and the final decision does not depend in a decisive way on individual elements taken in isolation, but rather on the internal relations among them.

A defining feature of deliberative tasks is the intrinsically noisy and partially contradictory nature of the labels. In realistic settings, cases that are apparently identical at the level of available information may have been assigned opposite labels at different times, reflecting subjective judgment, contextual effects, or incomplete knowledge. Under these conditions, a decision maker cannot rely on deterministic rules, but must instead construct a coherent and compromise-based evaluative framework grounded in previously observed cases.

Within this broad class, we are particularly interested in situations of \emph{hard deliberation}. By this term we denote deliberative tasks in which, in addition to the above features, the description of each decision element must be reduced to a dichotomic representation, even when this is not the original form of the input, in order to ensure the feasibility and effectiveness of the deliberative process.
In such cases, the informational content of the input is deliberately coarse-grained, and any meaningful decision must necessarily emerge from relational and collective properties rather than from fine-grained features.

\section{Outline of an Extreme Quantum Cognition Machine  for deliberative tasks}
\label{sec:QEL}

\begin{figure}
\includegraphics[width=1\textwidth]{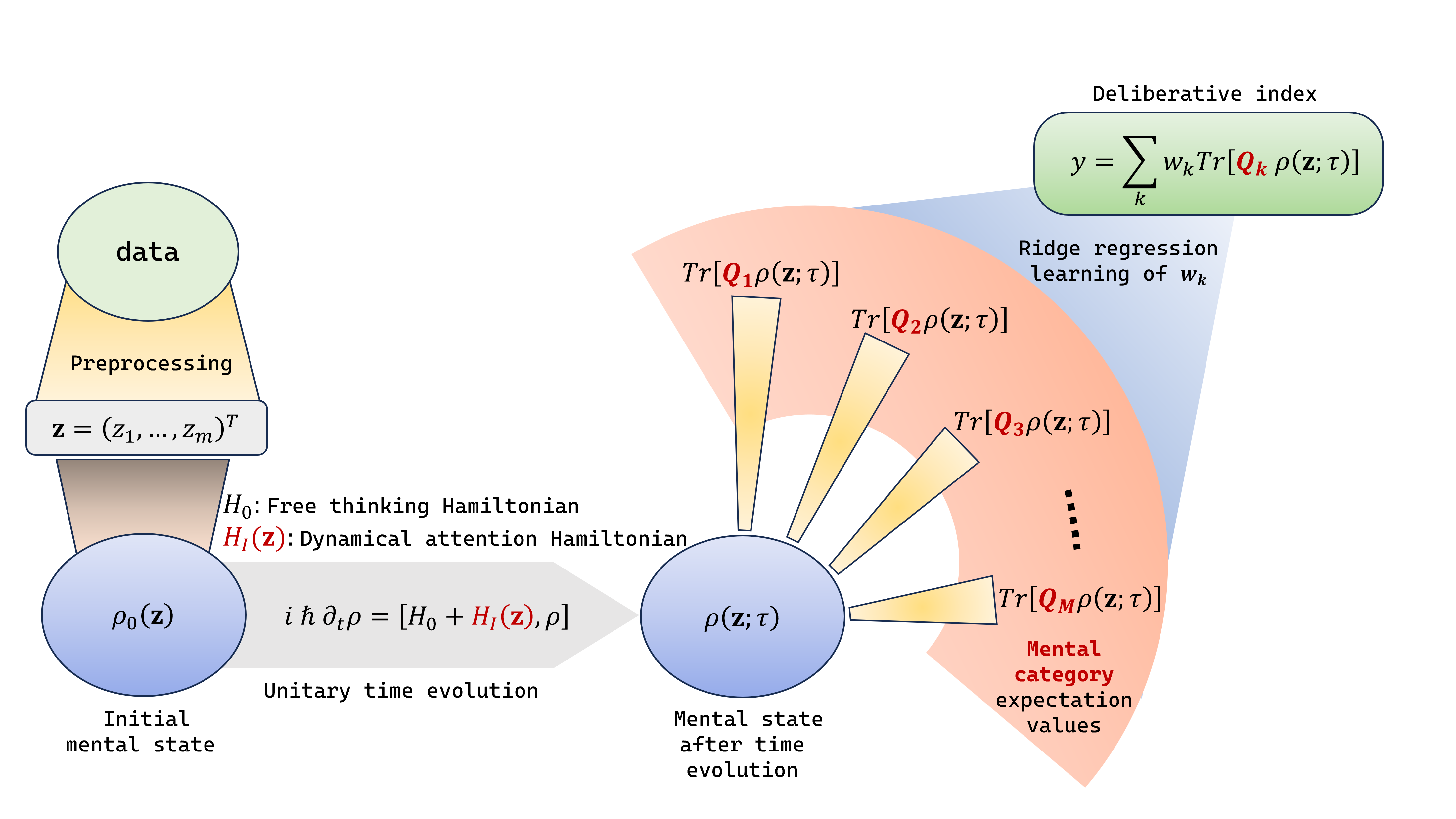}\\
\caption{Schematic representation of the Extreme Quantum Cognition Machine. Classical inputs $\boldsymbol{z}=(z_1,\dots,z_m)^{T}$, obtained after preprocessing of raw data, are mapped into a maximum-entropy density matrix $\rho_0(\boldsymbol{z})$, representing the initial mental state compatible with prescribed local expectation values. The state evolves unitarily under the Hamiltonian $H=H_0 + H_I(\boldsymbol{z})$, where $H_0$ models unguided (free-thinking) dynamics and $H_I$ encodes input-dependent dynamical attention, yielding $\rho(\boldsymbol{z};\tau)$. Expectation values of a fixed family of observables $\{Q_k\}_{k=1}^{M}$ define mental category features $f_k(\boldsymbol{z})=\mathrm{Tr}[Q_k \rho(\boldsymbol{z};\tau)]$. The final output (deliberative index) is obtained as a linear combination $y=\sum_k w_k f_k(\boldsymbol{z})$, where the weights $w_k$ are learned via ridge regression, while the quantum dynamical feature map remains fixed.}
\label{fig1}
\end{figure}

In this section, we provide the outline of an EQCM (see Fig. \ref{fig1}), which is specifically designed to operate as a quantum deliberator.\\
After suitable raw data preprocessing (see Sec. \ref{sec:encoding}), the input to the system is transformed into a classical vector $\boldsymbol{z} \in \{-\Delta,+\Delta\}^m$, representing the dichotomic values of the $m$ decision elements. The input vector $\boldsymbol{z}$ is used to construct an initial mental state, represented by a density matrix $\rho_0$ (see Sec. \ref{sec:initialization}). This state encodes the system’s first impression of the input and provides the starting point for the deliberative process. The state $\rho_0$ is then evolved unitarily for a time $\tau$ (see Sec. \ref{sec:evolution}) according to the von Neumann equation under a Hamiltonian decomposed, as usual in quantum mechanics, into free and interaction contributions, namely
\begin{equation}
H = H_0 + H_I ,
\end{equation}
with $[H_0,H_I]\neq0$. The resulting state is given by
\begin{equation}
\rho(\tau) = U(\tau)\,\rho_0\,U^\dagger(\tau), 
\end{equation}
with $U(\tau)=\exp(-\mathrm{i}H\tau)$,
and represents the outcome of a coherent quantum evolution driven by the internal dynamics and its interaction with the input.

Here, $H_0$ represents an input-independent component, corresponding to free and unconstrained internal dynamics, while $H_I$ describes the interaction with the input. From a dynamical perspective, this evolution can be interpreted as a coherent quantum walk in the space of internal states, where the input-dependent term $H_I$ selectively guides the exploration toward subspaces encoding the relational structure of the input. In this sense, $H_I$ plays a role analogous to attention mechanisms in contemporary artificial intelligence models \cite{vaswani2017attention}, by shaping the internal dynamics in response to the input.

After the evolution time $\tau$, the resulting state $\rho(\tau)$ is used to evaluate a deliberative observable 
\begin{equation}
    O_{\boldsymbol{w}} = \sum_k w_k Q_k ,
\end{equation}

where the operators $Q_k$ represent internal mental categories, and the real coefficients $w_k$ determine their relative importance (see Sec. \ref{sec:DelOp} for details about the spectral properties of $O_{\boldsymbol{w}}$). The continuous deliberative index is given by the expectation value
\begin{equation}
y = \mathrm{Tr}\!\left[\rho(\tau)\, O_{\boldsymbol{w}}\right].
\end{equation}

Learning in the EQCM framework consists in determining the weights $w_k$ from data using a linear training procedure, such as ridge regression (see Sec.~\ref{sec:redout}), while quantum dynamics remains fixed. As a result, learning in an EQCM does not modify the underlying quantum evolution, but produces a new deliberative operator $O_{\boldsymbol{w}}$, parametrized by the available mental categories and inferred from the training data. This operator represents the deliberative category effectively learned by the system from the corpus of past cases and encodes how internal evaluations are combined to produce the final decision.

\section{Algorithmic implementation}
\label{sec:ALG}

In this section, we provide a detailed description of the algorithmic components underlying the EQCM introduced in the previous section. The purpose is to make explicit how the abstract architecture outlined above is concretely instantiated by specifying the individual stages of the processing pipeline while keeping the overall structure modular.

\subsection{Encoding of symbolic strings into classical feature vectors}
\label{sec:encoding}

In a hard deliberation setting, the raw input data appear in the form of a vector
\[
\boldsymbol{s} = (s_1, s_2, \dots, s_m),
\]
where each decision element $s_i$ takes values in a finite and discrete alphabet $\mathcal{A}=\{a_1,a_2,\dots,a_{|\mathcal{A}|}\}$. This situation is rather general since, even when the original data present a different structure, appropriate preprocessing stages, such as discretization, thresholding or categorical encoding, can be applied to map the input to this form.

In hard deliberation problems, however, retaining the full resolution of each decision element is neither necessary nor desirable. Since the available information is intrinsically coarse-grained and the labels are noisy and partially contradictory, encoding each possible symbol independently would mainly inject noise into the model, increasing dimensionality without improving the quality of the deliberative inference. For this reason, we adopt a minimal encoding strategy in which each decision element is mapped onto a binary variable. This choice represents both the simplest and the most effective representation for hard deliberation tasks, striking an optimal balance between expressive power and robustness.

The construction of such a binary encoding is guided by basic principles of information theory \cite{shannon1948mathematical}. Given a discrete source emitting symbols with empirical probabilities $\{p(a)\}_{a\in\mathcal{A}}$, the self-information associated with a symbol $a$ is defined as
\[
I(a) = -\log p(a),
\]
while the Shannon entropy of the source is given by
\begin{equation}
\mathcal{H} = -\sum_{a\in\mathcal{A}} p(a)\log p(a).
\end{equation}
From this perspective, highly frequent symbols carry low self-information and form a largely uninformative background, whereas rare symbols correspond to high self-information events and represent informative deviations from typical patterns.

Based on this observation, we partition the alphabet $\mathcal{A}$ into two disjoint bins, $\mathcal{A}_{\mathrm{f}}$ and $\mathcal{A}_{\mathrm{r}}$, containing frequent and rare symbols respectively, chosen so as to be approximately equiprobable. The notion of frequent or rare symbols is inferred from the labeled training set and requires some care, as discussed in Sec.~\ref{sec:BENCH}.
This binning procedure effectively transforms the original symbolic source into a binary source with near-maximal entropy. As a consequence, information about the specific identity of individual symbols is deliberately discarded, and the resulting encoded strings no longer carry information about the absolute frequency of rare symbols.

In this representation, the informative content does not reside in the presence of individual rare symbols, but rather in the correlations, alternations, and collective patterns formed by symbols belonging to the two bins along the string. This shift of focus from symbol-level resolution to relational structure is a defining feature of hard deliberation, where meaningful decisions must emerge from global patterns rather than from isolated local features.

Within this binning scheme, each decision element answers a single binary question of the form: \emph{Is the decision element associated with an informative (i.e. low-probability) outcome?} The resulting dichotomic representation assigns a value $\pm \Delta$, with $\Delta \in \ ]0,1]$, to each element according to
\[
z_i =
\begin{cases}
+\Delta, & s_i \in \mathcal{A}_{\mathrm{r}}, \\
-\Delta, & s_i \in \mathcal{A}_{\mathrm{f}},
\end{cases}
\]
yielding a classical feature vector $\boldsymbol{z} \in \{-\Delta,+\Delta\}^m$ that constitutes the minimal representation required for the subsequent deliberative dynamics.

Importantly, this encoding admits a natural interpretation within the quantum cognition framework. The binary question associated with each decision element corresponds to a conceptual evaluation, represented in quantum cognition by an operator. In this perspective, the dichotomic value assigned to a decision element should be interpreted as the expectation of the corresponding operator over many hypothetical measurement instances. This interpretation aligns the classical preprocessing stage with the quantum formalism employed in the subsequent stages of the EQCM, where deliberative outcomes are consistently expressed in terms of expectation values.

\subsection{Quantum state initialization from classical inputs}
\label{sec:initialization}

The next step of the processing pipeline consists in mapping the classical feature vector
\[
\boldsymbol{z} = (z_1, z_2, \dots, z_m), \qquad z_k \in \{-\Delta,+\Delta\},
\]
into an initial quantum state $\rho_0$ describing the system’s mental state prior to the deliberative dynamics. Conceptually, this transformation is analogous to the operation known as \emph{fuzzyfication} in fuzzy inference systems \cite{KlirYuan1995}, where a crisp numerical input is mapped into a fuzzy set representing degrees of membership. 

Within the quantum cognition framework, each component $z_k$ of the input vector is interpreted as the effective outcome of a conceptual evaluation associated with a binary question, represented by an operator $O_k$. In the present context, $O_k$ corresponds to the question: \emph{Is the decision element associated with an informative (i.e.\ low-probability) outcome?} A projective measurement of $O_k$ would induce a probabilistic collapse of the mental state into one of its eigenstates. The quantity $z_k$, however, is not interpreted as the outcome of a single measurement, but rather as the expectation value of $O_k$ over many hypothetical repetitions of the same evaluative process under identical conditions. Accordingly, the initialization procedure is defined by the constraints
\[
z_k = \mathrm{Tr}\!\left[\rho_0\, O_k\right], \qquad k=1,\dots,m .
\]

Given the nature of the encoding described in Sec.~\ref{sec:encoding}, and without loss of generality, we assume that the operators $\{O_k\}$ are mutually commuting,
\[
[O_k, O_{k'}]=0 \qquad \forall\, k,k' ,
\]
so that they represent compatible evaluative variables. This allows us to choose a common eigenbasis and to identify the operators $O_k$ with Pauli operators acting on distinct degrees of freedom,
\[
O_k \equiv \sigma_z^{(k)} ,
\]
where $\sigma_z^{(k)}$ acts nontrivially only on the $k$-th subsystem. The commutativity of these operators plays a crucial role in ensuring that the encoding admits a simple and explicit representation (see Appendix \ref{AppA}).

The initial mental state $\rho_0$ is then determined as the quantum state that satisfies the expectation value constraints
\[
\mathrm{Tr}\!\left[\rho_0\, \sigma_z^{(k)}\right] = z_k ,
\]
while at the same time maximizing the von Neumann entropy
\begin{equation}   
S(\rho) = -\mathrm{Tr}\!\left(\rho \log \rho\right).
\end{equation}
It is important to emphasize that the von Neumann entropy quantifies the lack of information about the quantum state of a system, playing a role analogous to that of Shannon entropy, which measures the uncertainty associated with a classical information source. Maximizing $S(\rho)$ under the above constraints selects the least biased quantum state compatible with the available information.

The resulting constrained optimization problem admits a unique solution of Gibbs form (see Appendix \ref{AppA} for details). Owing to the commutativity of the operators $\sigma_z^{(k)}$, the solution factorizes as a tensor product of single-site density matrices,
\[
\rho_0 = \bigotimes_{k=1}^m \rho_k, \qquad
\rho_k = \frac{1}{2}\left(I + z_k\, \sigma_z\right).
\]
This explicit form of the encoding is made possible precisely by the compatibility of the evaluative operators, and ensures that no additional correlations or biases are introduced beyond those dictated by the classical input vector $\boldsymbol{z}$.

From a conceptual standpoint, this choice of $\rho_0$ provides a natural representation of the system’s first impression of the input data. The state is maximally unbiased except for the constraints imposed by the observed expectation values, reflecting a situation in which the system has incorporated the available information but has not yet developed higher-order correlations or contextual structure. Such correlations will instead emerge dynamically during the subsequent coherent evolution, as described in the following section.

\subsection{Quantum evolution and quantum-walk reservoir dynamics}
\label{sec:evolution}

The deliberative dynamics of an EQCM is governed by a Hamiltonian of the form
\[
H = H_0 + H_I ,
\]
where the two contributions play conceptually distinct and complementary roles. While the specific choice of these terms may in general depend on the task under consideration, one may reasonably expect that certain selection strategies exhibit a broad degree of validity and can be regarded as largely task agnostic.

We begin by discussing the input-independent component $H_0$. This term is intended to model what may be described, in cognitive terms, as free mental dynamics, i.e. unconstrained associations among internal categories, exploratory trajectories, and even paradoxical or seemingly irrelevant connections. Such dynamics encompass what is commonly referred to as imagination or free thought. Although a faithful microscopic model of these processes is clearly out of reach, insights from neuroscience suggest that healthy cognitive systems operate close to a regime of self-organized criticality \cite{beggs2003neuronal,chialvo2010emergent,shew2013functional}, in which the response function spans a wide range of scales, and the system exhibits maximal sensitivity to external stimuli. Working near such a critical regime allows the system to efficiently explore a rich internal state space and to flexibly integrate new information.

Within this perspective, we consider two complementary strategies for modeling $H_0$. The first consists in choosing $H_0$ as a random Hermitian matrix with real entries drawn from a Gaussian distribution with zero mean and fixed variance (i.e. random matrices belonging to the Gaussian Orthogonal Ensemble) \cite{haake1991quantum}. This choice generates a highly mixing and irregular quantum dynamics, providing a generic source of complex internal evolution. Such an approach naturally aligns with the philosophy of extreme learning and reservoir computing, where a fixed, randomly initialized dynamical system is used to generate a high-dimensional feature space \cite{havlivcek2019supervised}, and learning is confined to a linear readout layer. In this sense, a random $H_0$ can be viewed as the quantum counterpart of classical random reservoirs widely employed in machine learning.

The main limitation of this strategy lies in its implementability on current quantum hardware. While random Hamiltonians are conceptually simple and effective from a modeling standpoint, their faithful realization typically requires completely nonlocal interactions and fine-grained control, which are not yet readily available in near-term quantum devices. As a consequence, this approach is better suited to classical simulations or to future generations of quantum hardware with fewer architectural constraints.

A second more hardware-friendly strategy is to model $H_0$ using structured Hamiltonians, such as Ising-type models with random couplings and local fields, tuned to operate in the vicinity of a quantum critical point or in a chaotic regime \cite{Geller2022}. Systems near criticality naturally display long-range correlations and enhanced susceptibility, thereby capturing, at least qualitatively, key features associated with self-organized criticality and free associative dynamics. Importantly, Ising-like Hamiltonians with tunable parameters are directly compatible with several existing quantum computing platforms, making this approach particularly appealing from an experimental perspective. While the two strategies differ in their practical feasibility, it is reasonable to expect that, as quantum hardware matures, the distinction between them will become less relevant and both may eventually be implemented on physical quantum devices.

We now turn to the input-dependent interaction term $H_I$, whose role is to bias the free dynamics generated by $H_0$ in accordance with the specific input under consideration. As discussed earlier, $H_I$ acts as a form of dynamical attention, ensuring that the internal evolution remains informed by the structure of the input pattern. Although the detailed form of $H_I$ may be task dependent, a simple and effective choice capturing the essential features of this mechanism is given by
\begin{equation}
H_I = - g_1 \sum_k z_k\, \sigma_z^{(k)} 
      - g_2 \sum_{i>j} z_i z_j\, \sigma_z^{(i)} \sigma_z^{(j)} ,
\end{equation}
where $g_1$ and $g_2$ are non-negative coupling constants controlling the relative strength of local and pairwise contributions.

The physical interpretation of this interaction becomes transparent at a mean-field level. The first term ensures that, during the evolution, the sign of the expectation value $\langle \sigma_z^{(k)} \rangle$ tends to align with the sign of the corresponding input component $z_k$, thereby preserving information about individual decision elements. The second term enforces a similar alignment at the level of correlations, biasing the expectation values $\langle \sigma_z^{(i)} \sigma_z^{(j)} \rangle$ to reflect the sign of the products $z_i z_j$. In this way, the interaction Hamiltonian encodes not only the local structure of the input, but also its pairwise relational features.

As a result, the combined dynamics generated by $H_0$ and $H_I$ realizes a coherent quantum walk in which free exploration of the internal state space is continuously modulated by the structure of the input pattern. This mechanism allows the system to internalize correlations present in the data and to ensure that subsequent deliberative evaluations remain sensitive to the relational organization of the decision elements.

\subsection{Linear readout and training procedure}
\label{sec:redout}

The final stage of the EQCM consists in the linear readout and training procedure. At this level, all learning is confined to a linear model acting on features generated by the fixed quantum dynamics, while the Hamiltonian parameters and the evolution time remain unchanged.

Given the evolved state $\rho(\tau)$, the deliberative index is defined as the expectation value
\begin{equation}
    y = \mathrm{Tr}\!\left[\rho(\tau)\, O_{\boldsymbol{w}}\right]
  = \sum_k w_k\, \mathrm{Tr}\!\left[\rho(\tau)\, Q_k\right] \equiv \boldsymbol{w} \cdot \boldsymbol{x},
\end{equation}
where $O_{\boldsymbol{w}}=\sum_k w_k Q_k$ is the deliberative operator introduced in Sec.~\ref{sec:QEL}, and the quantities
\[
x_k = \mathrm{Tr}\!\left[\rho(\tau)\, Q_k\right]
\]
constitute the components of a classical feature vector $\boldsymbol{x}$ generated by the quantum evolution.

Thus, each input pattern, labeled with a dichotomic target label $t \in \{t_{+},t_{-}\}$ representing the desired deliberative outcome,  produces a feature vector $\boldsymbol{x}$. Collecting the feature vectors and the expected deliberative outcomes over the available dataset defines a training set $\{(\boldsymbol{x}^{(\mu)}, t^{(\mu)})\}_{\mu=1}^{N}$, which is used to determine the optimal weights $\boldsymbol{w}$ by minimizing a regularized quadratic loss function. Specifically, the weights are obtained via ridge regression by solving a convex optimization problem.

To make the optimization procedure explicit, we introduce the training data matrix $X$ and the target vector $\mathbf{t}$. 
Given $N$ training samples with feature vectors 
$\boldsymbol{x}^{(\mu)} \in \mathbb{R}^M$ and target labels 
$t^{(\mu)} \in \mathbb{R} $, we define the design matrix 
$X \in \mathbb{R}^{N \times M}$ with entries
\begin{equation}
X_{\mu i} = x^{(\mu)}_i ,
\end{equation}
and the target vector $\mathbf{t} = (t^{(1)}, \dots, t^{(N)})^{\mathsf T}$.

The ridge regression problem \cite{hoerl1970ridge}, written in matrix form, amounts to minimizing the regularized quadratic loss
\begin{equation}
\mathcal{L}(\boldsymbol{w}) 
= \| X\boldsymbol{w} - \mathbf{t} \|_2^2 
+ \lambda \|\boldsymbol{w}\|_2^2 ,
\end{equation}
where $\lambda > 0$ controls the trade-off between data fitting and model complexity. The regularization strength $\lambda$ plays the role of a hyperparameter of the model, along with the evolution time $\tau$, the coupling constants entering $H_I$, and the fixed parameters defining the Hamiltonian dynamics.
Setting the gradient of $\mathcal{L}(\boldsymbol{w})$ with respect to $\boldsymbol{w}$ equal to zero yields the normal equations
\begin{equation}
(X^{\mathsf T} X + \lambda I)\boldsymbol{w} 
= X^{\mathsf T}\mathbf{t},
\end{equation}
whose unique solution is
\begin{equation}
\boldsymbol{w} 
= (X^{\mathsf T} X + \lambda I)^{-1} X^{\mathsf T}\mathbf{t}.
\end{equation}

Because the loss function is strictly convex for $\lambda>0$, the latter represents the unique global minimum of the optimization problem. This renders the training procedure computationally efficient, robust to noise and label fluctuations, and intrinsically insensitive to initialization, in contrast with iterative and non-convex learning schemes commonly employed in deep neural networks.

From a computational perspective, this linear training strategy scales favorably with the size of the dataset and the feature dimension, and requires significantly reduced training times compared to gradient-based methods. These characteristics make the architecture particularly well suited for scenarios in which rapid retraining, online updates, or integration within agent-based learning loops are required.

An important aspect of the learning outcome concerns the interpretation of the continuous index $y$. In deliberative tasks with noisy and partially contradictory labels, the nominal targets $t_{\pm}$ are typically followed with a nonzero dispersion. After successful training, one expects the values of $y$ evaluated on inputs labeled by $t_{+}$ ($t_{-}$) to be distributed around $t_{+}$ ($t_{-}$), with a variance reflecting the intrinsic ambiguity and inconsistency of the deliberative process. When the model captures the relevant structure of the task, the two resulting distributions exhibit negligible overlap. Conversely, a significant overlap signals inadequate learning, which may arise either from a suboptimal encoding of the inputs or from intrinsic difficulties associated with the task itself. A more detailed analysis of these effects is deferred to the discussion of the numerical results (Sec. \ref{sec:BENCH} and \ref{sec:HF}).

Because the output of the model is a real-valued deliberative index, a sharp decision requires the adoption of an explicit decision strategy. In this work, we consider the simple choice of assigning the predicted label according to the sign of $y$, while noting that alternative and potentially more sophisticated decision rules may also be employed. In this framework, model performance is naturally quantified using classification metrics derived from the confusion matrix, including false positives, false negatives, accuracy, and balanced accuracy. These metrics will be employed throughout the remainder of this work.

\section{Linguistic benchmark and performance analysis}
\label{sec:BENCH}

In this section we assess the proposed EQCM architecture on two symbolic classification tasks designed to exemplify the notion of hard deliberation introduced in Sec.~\ref{sec:QC}. Although both tasks are formulated in a linguistic setting, they are not intended as benchmarks for language recognition per se. Rather, they serve as controlled and interpretable testbeds for deliberative decision making under conditions of reduced local information, noisy labels, and intrinsically relational structure.

In both tasks, the input consists of fixed-length symbolic strings, and the objective is to assign a binary label to each input. Crucially, the decision cannot be inferred from the presence of specific symbols at individual positions, but must instead emerge from global patterns and correlations across the entire string. This feature places the tasks squarely within the class of hard deliberation problems.

The first task (\textit{Task 1}) consists in discriminating Italian words from random seven-letter strings. While seemingly simpler, this task remains deliberative in nature due to the severe reduction of the input representation and the absence of explicit symbolic information at the local level. Successful classification therefore requires the identification of global structural patterns rather than the detection of isolated symbols.\\
The second task (\textit{Task 2}) consists in discriminating seven-letter words drawn from Italian and English dictionaries. This task represents a form of fine-grained deliberation, as the two classes share comparable statistical properties and overlapping local features, while differing subtly in their global relational structure.

In both cases, the original symbolic information is deliberately compressed through a hard encoding strategy, enforcing a dichotomic representation of each decision element. As a consequence, the post-processed input no longer carries explicit information about individual symbols, and successful classification must rely on the collective organization of the representation and on the correlations preserved by the encoding. Importantly, distinct symbolic inputs belonging to opposite classes may become indistinguishable after encoding, introducing an intrinsic form of label ambiguity even in the absence of externally imposed noise. This reduction is a defining feature of hard deliberation, where the system must operate under conditions of information scarcity and partial ambiguity.

Throughout this section, the quantum reservoir dynamics is generated using a task-agnostic internal Hamiltonian \(H_0\), drawn once from the Gaussian Orthogonal Ensemble (GOE) and kept fixed throughout the analysis, possibly accompanied by the dynamical attention term \(H_I\). This choice reflects the extreme learning philosophy adopted in this work, whereby the internal dynamics is fixed and affected by a random contribution, while learning is confined to a linear readout. We consider a single random realization of \(H_0\) since the large number of random matrix elements is expected to produce self-averaging effects, which should make the observed performance rather insensitive to the specific realization. This supports the view that the deliberative capabilities do not rely on task-specific fine-tuning of the internal Hamiltonian, but rather on the interplay between the encoded input, quantum evolution, and learned readout.

Taken together, these tasks provide paradigmatic examples of hard deliberation problems that transcend their specific linguistic realization. Indeed, linguistic strings serve only to generate dichotomic input vectors with nontrivial statistical correlations, which constitute the actual inputs to the EQCM architecture. Similar correlated representations can arise in a wide range of contexts, including symbolic inference, anomaly detection, biological sequence analysis, and diagnostic tasks, where decisions must be formed from globally structured yet locally impoverished data. The analysis presented below is therefore intended to highlight general mechanisms underlying quantum deliberative processing, independently of the specific origin of the input data.

\subsection{Label-aware and task-informed encoding strategies}

Before analyzing the two deliberative tasks in detail, it is useful to further clarify the notion of rare and frequent symbols introduced in Sec.~\ref{sec:encoding}, and to discuss how this notion should be operationally defined in the presence of labeled data. While the discussion is of general relevance, it naturally finds a clear and intuitive illustration in the concrete tasks considered here.

The maximum-entropy two-bin encoding strategy requires an estimate of the empirical frequencies of the symbols in the underlying alphabet. In practice, this estimation can be carried out following two distinct approaches. A first possibility is to infer symbol frequencies from the entire training set, independently of the associated labels. A second possibility is to restrict the frequency analysis to a subset of the training data with a fixed deliberative label. As we argue below, the choice between these strategies plays a crucial role in determining the effectiveness of the encoding and, consequently, the overall deliberative performance.

If the symbolic strings to be classified are generated by a single information source characterized by a well-defined emission probability for each symbol, then acquiring statistics over the entire training set is a consistent and natural choice. In this case, the notion of rarity is globally meaningful, and the resulting two-bin encoding provides a faithful coarse-grained representation of the underlying source.

The situation changes qualitatively when the data are produced by two distinct sources with different emission statistics, as is the case for strings drawn from different natural languages or from structured versus random processes. In such scenarios, pooling together samples from both classes to estimate symbol frequencies effectively reconstructs a fictitious source, whose emission probabilities are a distorted version of the original ones. The resulting notion of rarity no longer reflects the statistics of either source and can lead to a severe degradation of the informative content retained by the encoding.

To avoid this issue, the frequency analysis must be performed in a label-aware manner, restricting the estimation to training samples associated with a fixed deliberative label. This procedure preserves the statistical structure of one of the underlying sources and allows the system to discriminate the alternative class by contrast. Importantly, either label can be chosen as the reference for the encoding, leading to two inequivalent label-aware representations. In general, these choices do not yield identical performance, and one of them typically proves more effective.

This asymmetry admits a natural interpretation within a deliberative framework. A label-aware encoding can be viewed as being constructed by an implicit expert that is highly familiar with one class of patterns and evaluates other inputs relative to this internal reference. Such an expert-based representation is a common feature of deliberative processes, where decisions are often formed by comparison with a well-established mental scenario rather than by symmetric treatment of all alternatives.

The considerations above are not specific to the linguistic examples discussed here, but reflect a general feature of hard deliberation tasks involving multiple information sources. Whenever labels correspond to distinct generative processes, a label-aware definition of rarity is necessary to prevent the loss of discriminative information induced by excessive coarse graining.

In addition to label-aware encodings, one may also consider task-informed encoding strategies, where the coarse graining of the alphabet is guided by structural knowledge about the nature of the task itself. In the present linguistic setting, a natural example is the consonant–vowel encoding, which partitions letters according to their phonological role. From a linguistic perspective, consonants and vowels play distinct functional roles in syllable structure and lexical organization, and their distribution is shaped by articulatory constraints, perceptual discriminability, and efficiency principles governing human communication. For instance, frequent words tend to be shorter, and phonotactic regularities constrain permissible sound sequences in ways that reflect both production and recognition constraints \cite{zipf1949human,nespor2003different,mazzarisi2021maximal}.

Such a task-informed encoding incorporates prior structural insight into the generative process underlying the data, and therefore may enhance discriminative performance when this structural knowledge is relevant. Importantly, the consonant–vowel partition can also be recovered, in principle, within the maximum-entropy two-bin framework described above, when the empirical statistics of the labeled training set reflect these deeper phonological regularities. In this sense, the two-bin label-aware procedure provides a data-driven mechanism capable of reproducing meaningful structural partitions even in the absence of explicit prior knowledge. 

More generally, whenever domain-specific structural information is available, incorporating it at the encoding stage can improve performance. Conversely, when such knowledge is not directly accessible, the label-aware maximum-entropy binning offers a principled and model-agnostic alternative for extracting latent structural distinctions from data. The interplay between these two perspectives will be illustrated concretely in the tasks analyzed below.

\subsection{Task 1: Deliberative discrimination between Italian words and random strings}

For Task 1, we consider the problem of discriminating Italian seven-letter words from random seven-letter strings. Inputs drawn from the Italian dictionary are assigned label $t_+=+0.5$, whereas randomly generated strings are assigned label $t_-=-0.5$. The training set is balanced and consists of 150 samples per class; the test set contains 40 samples per class.

Italian lexical data are obtained from a standard computational dictionary and filtered according to the following criteria: fixed length equal to seven characters, absence of hyphens and apostrophes, and lowercase initial character. This yields a pool of 10912 admissible Italian seven-letter words. Random strings are generated by sampling uniformly and independently from the 26-letter lowercase Latin alphabet, thereby ensuring the absence of lexical or phonological structure beyond trivial uniform statistics.

All words are represented over a fixed discrete alphabet
\[
\mathcal{A} = \{a, \grave{a}, b, c, d, e, \grave{e}, \acute{e}, f, g, h, i, \grave{i}, j, k, l, m, n, o, \grave{o}, p, q, r, s, t, u, \grave{u}, v, w, x, y, z\},
\]
where accented vowels are treated as distinct symbols when present in the source data.

In the consonant–vowel encoding adopted here, the alphabet is partitioned into two phonological classes. The vowel set is defined as
\[
\mathcal{V} = \{a, \grave{a}, e, \grave{e}, \acute{e}, i, \grave{i}, o, \grave{o}, u, \grave{u}, j, w\},
\]
while the consonant set is $\mathcal{C} = \mathcal{A} \setminus \mathcal{V}$. 

The inclusion of the semivowels $j$ and $w$ in $\mathcal{V}$ reflects their phonological role. In many linguistic contexts, these symbols behave as glides and share articulatory and acoustic properties with vowels, contributing to syllabic structure and vowel-like transitions. Treating them within the vowel class is therefore consistent with a coarse-grained phonological partition aligned with articulatory similarity.

Before encoding, each word undergoes a tokenization step, defined as the ordered decomposition of the string into its constituent characters:
\[
s = (s_1, s_2, \dots, s_7).
\]
For instance, the Italian word \texttt{abbagli} is tokenized as
\[
(\texttt{a},\texttt{b},\texttt{b},\texttt{a},\texttt{g},\texttt{l},\texttt{i}).
\]

Each letter plays the role of a decision element, which is then mapped to a dichotomic value according to
\[
z_k =
\begin{cases}
+1 & \text{if } s_k \in \mathcal{C}, \\
-1 & \text{if } s_k \in \mathcal{V}.
\end{cases}
\]

Applying this rule to \texttt{abbagli} yields
\[
\boldsymbol{z} = (-1,+1,+1,-1,+1,+1,-1),
\]
since $a,i \in \mathcal{V}$ and $b,g,l \in \mathcal{C}$.

The consonant–vowel partition underlying this mapping is not arbitrary. In Italian, vowels are statistically more frequent and form a relatively uniform background component of words, while consonants tend to carry stronger discriminative structure through combinatorial constraints and positional correlations. From an information-theoretic perspective, vowels therefore behave as higher-probability (lower self-information) symbols, whereas consonants act as relatively rarer and more informative events.

The resulting binary vectors $\boldsymbol{z} \in \{-1,+1\}^7$ constitute the classical inputs to the quantum initialization stage described in Sec.~\ref{sec:initialization}. In this task-informed encoding, phonological structure provides an inductive bias aligned with statistical regularities of the language, while remaining compatible with the general two-bin maximum-entropy framework introduced earlier.

Once the binary encoding $\boldsymbol{z} \in \{-1,+1\}^7$ is obtained, the input is mapped to an initial quantum state $\rho_0(z)$ and evolved unitarily up to time $\tau$, yielding $\rho(\tau; z)$. From this state we construct a real-valued feature vector whose components are expectation values of a fixed set of Hermitian operators $\{Q_k\}$.

In the present implementation, the operator family includes local single-qubit observables and short-range correlators of increasing order. More precisely, the feature vector contains:

(i) all single-site expectation values
\[
\langle \sigma_x^{(k)} \rangle,\quad
\langle \sigma_y^{(k)} \rangle,\quad
\langle \sigma_z^{(k)} \rangle,
\qquad k = 1,\dots,7;
\]

(ii) nearest-neighbour two-body correlators
\[
\langle \sigma_\alpha^{(k)} \sigma_\alpha^{(k+1)} \rangle,
\qquad \alpha \in \{x,y,z\},\quad k = 1,\dots,6;
\]

(iii) higher-order longitudinal correlators of the form
\[
\langle \sigma_z^{(k)} \sigma_z^{(k+1)} \cdots \sigma_z^{(k+r)} \rangle,
\]
with $r = 2,\dots,6$ and $k$ chosen so that the support remains within the seven-qubit register;

(iv) the expectation value of the identity operator on the full Hilbert space,
\[
\langle \mathcal{I} \rangle = \mathrm{Tr}\big[\rho(\tau; z)\,\mathcal{I}\big] = 1,
\]
which is included as the final component of the feature vector, thus adding a constant unit feature that plays the role of a bias term in the linear readout.

Collecting all these expectation values defines a feature vector
\[
\boldsymbol{x}(z) = \big( \langle Q_1 \rangle, \dots, \langle Q_M \rangle \big) \in \mathbb{R}^{M},
\]
where each component is given by
\[
\langle Q_k \rangle = \mathrm{Tr}\big[ \rho(\tau; z)\, Q_k \big].
\]
Once the feature vector $\boldsymbol{x}(z)$ is calculated for each input in the training set, the weights $\boldsymbol{w}$ are obtained by ridge regression, as detailed in Sec.~\ref{sec:redout}. 
In this way, learning selects the optimal linear combination of internal categories, $Q_k$, defining an effective deliberative operator $O_{\boldsymbol{w}}$,
whose expectation value yields the continuous deliberative index $y=\boldsymbol{w} \cdot \boldsymbol{x}$.\\
A sharp decision is obtained from the continuous index $y$ through a simple thresholding strategy, assigning a string to the Italian class if $y>0$ and to the random class if $y<0$, i.e. according to the value of $\mathrm{sign}(y)$. While more refined decision criteria could in principle be adopted, their discussion is not essential for the purposes of the present work.\\
Once the weights are identified and the system is trained, its performance is evaluated on both the training and test sets using classification metrics derived from the confusion matrix.\\
For a binary classification problem, performance is naturally assessed in terms of the confusion matrix \cite{pearson1904theory,miller1955analysis}. Denoting by the \emph{positive} class the Italian words and by the \emph{negative} class the random strings, each prediction falls into one of four categories: true positives (TP), false negatives (FN), false positives (FP), and true negatives (TN). Here TP counts Italian words correctly classified as Italian, FN Italian words misclassified as random, FP random strings incorrectly assigned to the Italian class, and TN random strings correctly classified as random.

From these quantities we derive the performance metrics used throughout the present work (see Appendix \ref{AppB} for details). The \emph{accuracy} is defined as
\begin{equation}
\mathrm{Accuracy} =
\frac{\mathrm{TP} + \mathrm{TN}}
{\mathrm{TP} + \mathrm{TN} + \mathrm{FP} + \mathrm{FN}},
\end{equation}
and measures the overall fraction of correctly classified samples. The \emph{balanced accuracy} is given by
\begin{equation}
\mathrm{BA} =
\frac{1}{2}
\left(
\frac{\mathrm{TP}}{\mathrm{TP} + \mathrm{FN}}
+
\frac{\mathrm{TN}}{\mathrm{TN} + \mathrm{FP}}
\right),
\end{equation}
and averages the recall over the two classes, providing a more informative metric when class distributions are unbalanced.

In addition, we consider class--conditional precision measures. For the Italian class,
\begin{equation}
\mathrm{Precision}_{\mathrm{IT}} =
\frac{\mathrm{TP}}{\mathrm{TP} + \mathrm{FP}},
\end{equation}
while for the random class,
\begin{equation}
\mathrm{Precision}_{\mathrm{RND}} =
\frac{\mathrm{TN}}{\mathrm{TN} + \mathrm{FN}}.
\end{equation}

These metrics provide complementary information on the reliability of positive and negative predictions, respectively. In what follows, model performance on both training and test sets will be summarized in terms of these quantities, allowing consistent comparison across encoding strategies and dynamical parameters.

\begin{figure}
\includegraphics[width=1\textwidth]{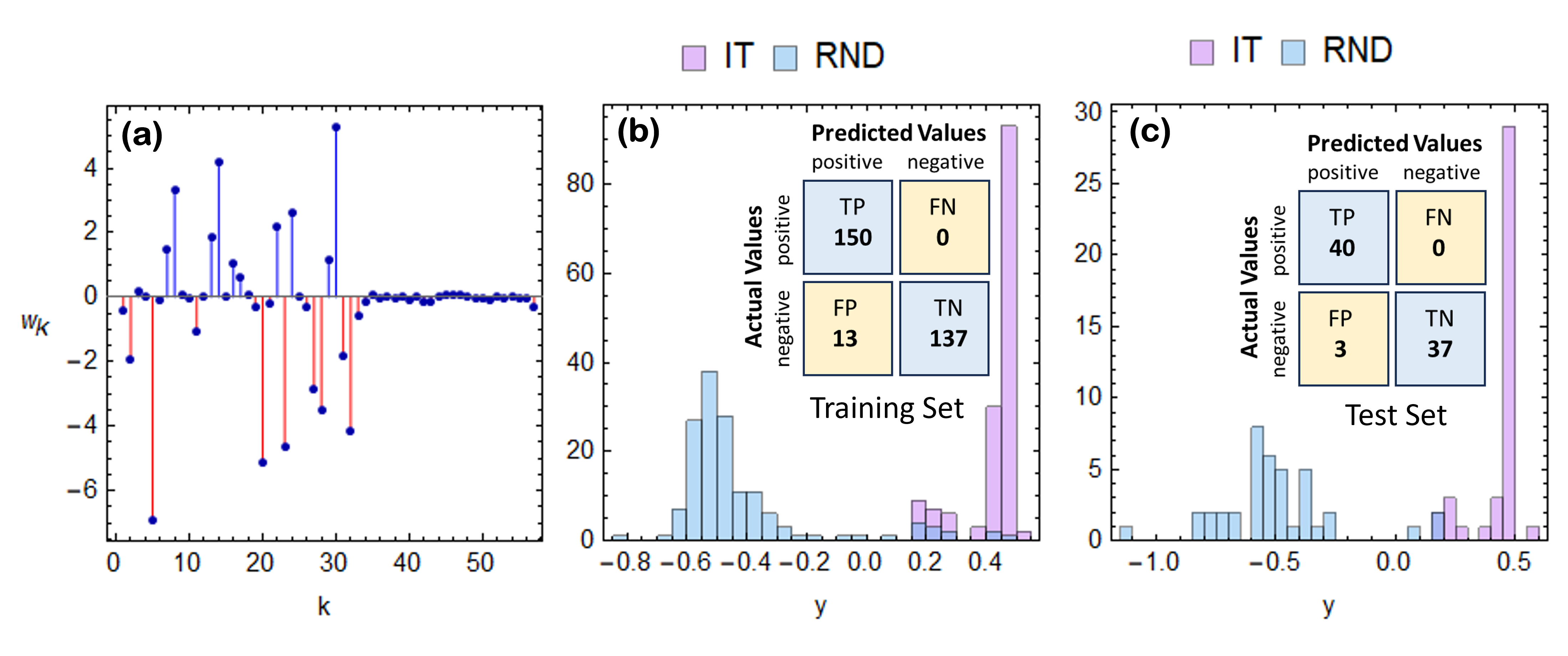}\\
\includegraphics[width=1\textwidth]{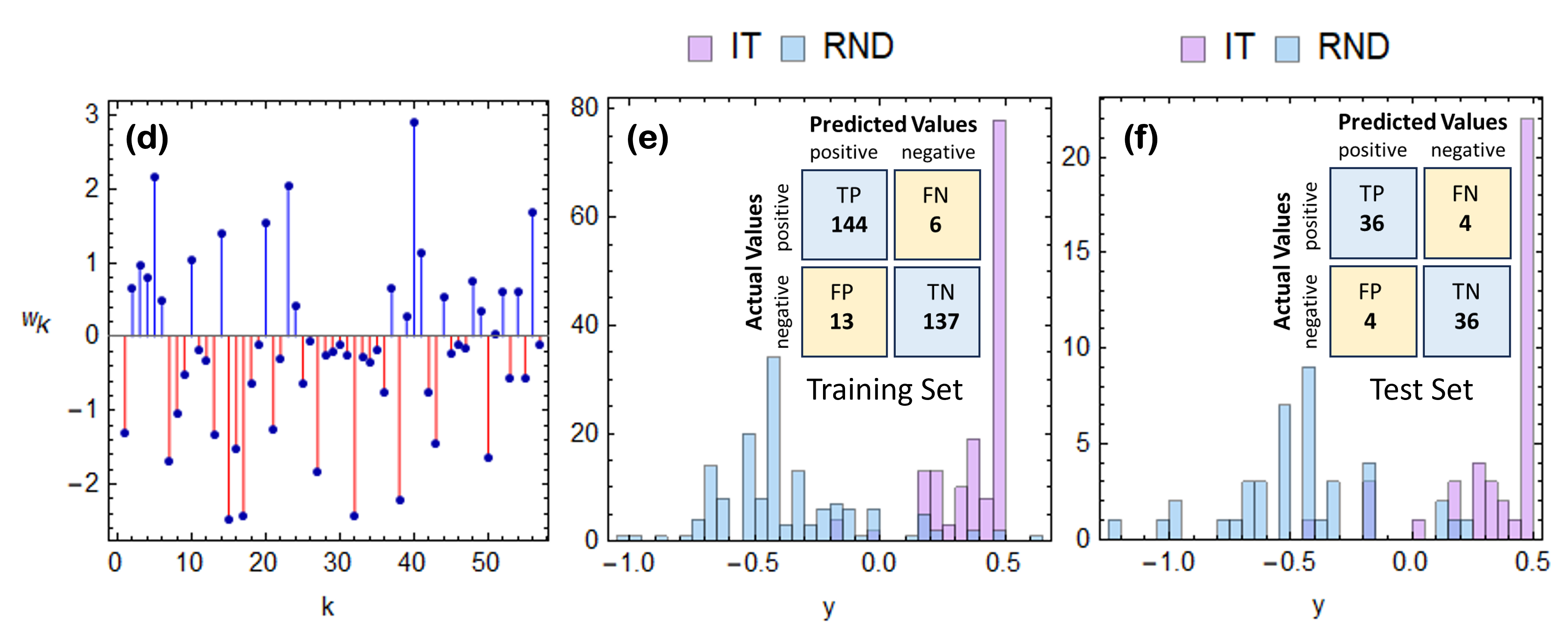}
\caption{Performance of the EQCM architecture on Task 1 (Italian vs random strings) with and without dynamical attention.\\ 
Panels (a)–(c) correspond to the setting 
$\sigma = 0.1$, $\tau = 10$, $\lambda = 2\times 10^{-3}$, 
$g_1 = 0.1$, $g_2 = 0.4$, where $\sigma$ is the variance of the real-valued GOE Hamiltonian $H_0$ (zero mean), $\tau$ the dimensionless evolution time, and $g_{1,2}$ the coupling strengths controlling the interaction Hamiltonian $H_I$. In this regime dynamical attention is active. 
Panel (a) shows the learned weights $w_k$: only categories involving single-qubit observables and nearest-neighbour two-qubit correlators acquire significant amplitudes, indicating that attention effectively restricts learning to local two-letter structures. 
Panels (b) and (c) display the distribution of the continuous deliberative index $y$ for the training and test sets, respectively, with the corresponding confusion matrices shown as insets. For the training set, accuracy and balanced accuracy are both $0.9567$; for the test set, accuracy and balanced accuracy are $0.9625$, confirming excellent generalization.\\
Panels (d)–(f) are obtained with identical hyperparameters but with $g_1 = g_2 = 0$, i.e. with attention switched off. In this case, (d) shows that weights are distributed over a broader set of internal categories, including higher-order correlators. The corresponding histograms (e) and (f) reveal a moderate degradation of performance (training accuracy $0.9367$, test accuracy $0.9000$), indicating that dynamical attention enhances selectivity and improves generalization by biasing the dynamics toward task-relevant local structures.}
\label{fig2}
\end{figure}

The results for Task~1 are summarized in Fig.~\ref{fig2}. We begin with the regime in which dynamical attention is active ($g_1 \neq 0$, $g_2 \neq 0$). In this setting, the learned weights exhibit a clear structural selectivity (see panel (a)). Indeed, significant amplitudes concentrate on single-site observables and nearest-neighbour two-qubit correlators, while higher-order longitudinal correlators remain suppressed. This indicates that the interaction term $H_I$ effectively biases the dynamics toward local two-letter structures, reducing the effective hypothesis space explored by the linear readout.

This structural sparsification translates into a clear statistical separation of the continuous deliberative index $y$ for the two classes (see panels (b) and (c)). On the training set (panel (b)), the model achieves $\mathrm{Accuracy} = 0.9567$ and $\mathrm{BA} = 0.9567$, with class-conditional precisions $\mathrm{Precision}_{\mathrm{IT}} = 0.9202$ and $\mathrm{Precision}_{\mathrm{RND}} = 1.000$. Importantly, the performance remains stable on unseen data (panel (c)), with test balanced accuracy $\mathrm{BA} = 0.9625$, indicating that the learned operator $O_w$ captures genuinely discriminative structure rather than overfitting the training set.

When the attention mechanism is switched off ($g_1 = g_2 = 0$), while keeping all other hyperparameters unchanged, the learned weights spread across a broader set of internal categories (see panel (d)), including higher-order correlators. This broader activation of features is accompanied by a mild but systematic degradation of performance, as reflected by a reduction of the training accuracy to 0.9367 and of the test accuracy to 0.9000.
Moreover, the increased overlap between the class-conditional distributions of $y$ (see panels (e) and (f)) reflects a reduced discriminative margin.

These results show that dynamical attention acts as an inductive bias at the level of the quantum evolution, guiding the feature embedding toward task-relevant local correlations. This bias improves both selectivity and generalization, as quantified by the metrics derived by the confusion matrix. Thus, the proposed EQCM architecture is capable of discriminating Italian words from random strings with high accuracy and stable generalization. 

A distinctive feature of the model is the emergence of a continuous deliberative index $y$ whose class-conditional distributions concentrate around the target values with different variances. In the label-aware setting (equivalent to the task-aware setting considered for Task 1), where the notion of rarity is calibrated with respect to one class (i.e. the set of Italian words), the corresponding distribution exhibits reduced dispersion, while the complementary class (i.e. the set of random strings) is recognized primarily by contrast, leading to a broader spread. This asymmetry reflects the underlying deliberative mechanism rather than an artifact of the linear readout.

Importantly, with dynamical attention active, the model produces no false negatives on either the training or the test set, meaning that Italian words are never misclassified as random strings. Misclassifications occur only in the opposite direction, where a small fraction of random strings are classified as Italian. This asymmetric error structure is consistent with the task-informed encoding and indicates a conservative decision boundary biased toward preserving the structured class. When attention is switched off, this property is lost, and errors become more symmetrically distributed.

These observations collectively support the interpretation of dynamical attention as a physically grounded mechanism that enhances selectivity, stabilizes the discriminative index, and improves generalization in hard deliberative tasks.

\subsection{Task 2: Deliberative discrimination between Italian and English words}

In the second task, the architecture is required to discriminate between Italian and English seven-letter words extracted from the respective computational dictionaries under the same filtering criteria adopted in Task~1 (fixed length equal to seven characters, lowercase initial letter, no hyphens or apostrophes). The resulting pools contain $10\,912$ admissible Italian words and $11\,968$ English words. From these, we construct balanced datasets consisting of $150$ samples per class for training ($300$ total inputs) and $40$ samples per class for testing ($80$ total inputs), as done in Task~1.

In contrast to Task~1, both classes now possess genuine lexical and phonological structure and differ only in their statistical organization. The problem is therefore intrinsically more demanding, as discrimination must rely on subtle structural regularities rather than on the mere presence or absence of linguistic coherence.

We compare two encoding strategies.

The first strategy coincides with the consonant--vowel partition introduced in Task~1. The alphabet $\mathcal{A}$ is divided into a vowel set $\mathcal{V}$ (including accented vowels and semivowels) and its complement $\mathcal{C} = \mathcal{A} \setminus \mathcal{V}$. Each tokenized word is mapped to a binary string $z \in \{-1,+1\}^7$ according to the same rule as before. This encoding is task-aware and reflects a coarse phonological prior.

The second strategy is a label-aware encoding based on two maximum-entropy buckets constructed from the empirical letter frequencies of the Italian training subset only. The encoding is obtained by collecting in the set $\mathcal{W}_{\mathrm{IT}}^{\mathrm{train}}$ the Italian words in the training set. After tokenization into individual characters, the empirical frequency distribution $p(\ell)$ over $\ell \in \mathcal{A}$ is obtained. The letters are sorted in decreasing order of frequency, and a split index is selected so that the cumulative probability of the first group is as close as possible to $1/2$, thereby producing two buckets with approximately equal total probability. This construction maximizes the Shannon entropy \cite{shannon1948mathematical} of the induced binary partition under the constraint of label awareness.

In practice, orthographic variants corresponding to the same phoneme, when present in $\mathcal{W}_{\mathrm{IT}}^{\mathrm{train}}$ (e.g., $o$ and $\grave{o}$, $a$ and $\grave{a}$), are merged prior to the entropy-based split, so as to avoid artificial fragmentation of the empirical statistics. After aggregation, for instance, the effective frequency of the letter $o$ is obtained as
\[
p(o) + p(\grave{o}) \simeq 0.06095 + 0.00952 = 0.07048,
\]
while the empirical frequency of $n$ is given by
\[
p(n) \simeq 0.06857.
\]
Since $p(o) + p(\grave{o}) > p(n)$, the letter $o$ (together with $\grave{o}$) precedes $n$ in the frequency-based ranking. As a consequence, we get the high-frequency bucket 
\[
\mathcal{A}_\mathrm{f}=\{a, \grave{a}, i, t, e, o, \grave{o}\},
\]
with empirical probability $p(\mathcal{A}_{\mathrm{f}}) \approx 0.52$, while the rare-letter bucket $\mathcal{A}_{\mathrm{r}}=\mathcal{A} \setminus \mathcal{A}_{\mathrm{f}}$ carries the complementary probability. We deliberately do not include accented variants of $i$ and $e$ in $\mathcal{A}_{\mathrm{f}}$, as they are absent from the Italian training subset ( $\mathcal{W}_{\mathrm{IT}}^{\mathrm{train}}$) used to estimate the empirical frequencies. Although, on linguistic grounds, such letters could be grouped with their unaccented counterparts, analogously to the aggregation performed for $o$ and $\grave{o}$, we retain the purely data-driven partition of the alphabet $\mathcal{A}$ in order to assess the robustness of the architecture under encodings inferred exclusively from finite-sample statistics, without heuristic corrections beyond entropy maximization.

Interestingly, $\mathcal{A}_\mathrm{f}$ contains almost all principal vowels (with the exception of $u$), together with the consonant $t$, reflecting the dominant role of vocalic structure and highly recurrent consonantal patterns in the Italian training subset. In this sense, the resulting partition can be viewed as a data-driven approximation of the consonant-vowel encoding introduced in Task~1. Once the partition $\mathcal{A}=\mathcal{A}_{\mathrm{f}} \cup \mathcal{A}_{\mathrm{r}}$ has been defined, each letter in a word is then encoded by assigning dichotomic values according to membership in the frequent bucket versus its complement, and the deliberative pipeline proceeds exactly as in Task~1.

In principle, one could construct analogous buckets using the English training subset or combine frequency information from both classes. We have explored these alternatives, which, however, present low performance. Indeed, since the two classes are generated from sources with distinct letter-frequency statistics, the encoding must be constructed from a label-homogeneous training subset. Combining frequencies from both classes produces a blurred empirical distribution in which class-specific statistical imbalances are partially degraded, thereby reducing discriminative information. Consistently with the inductive-bias perspective developed in the previous sections, we calibrate the encoding with respect to a reference class (i.e., that of the Italian words), so that discrimination emerges through contrast against a competing structure. We have further verified that the encoding calibrated on Italian statistics yields better performance than the one derived from the English subset, reflecting a closer alignment with language-agnostic phonological constraints and with general motor-articulatory principles underlying speech production.

\begin{figure}
\includegraphics[width=1\textwidth]{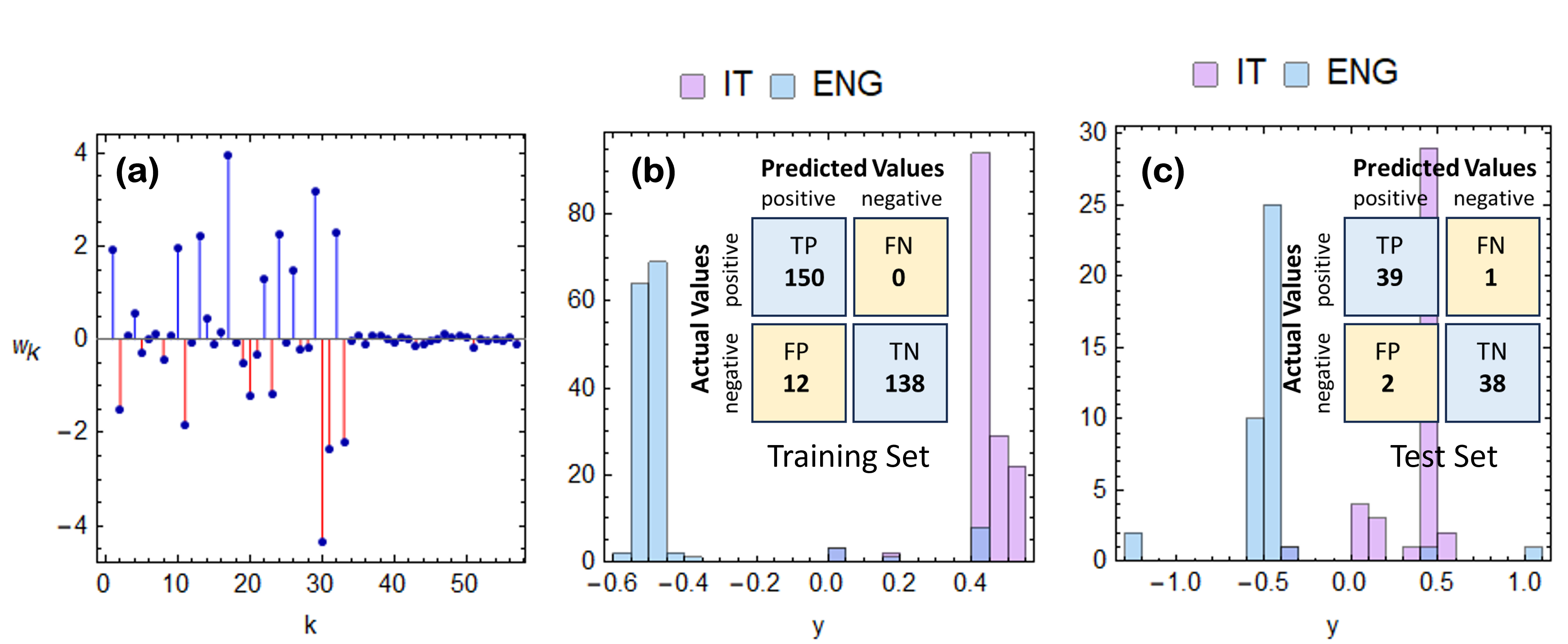}\\
\includegraphics[width=1\textwidth]{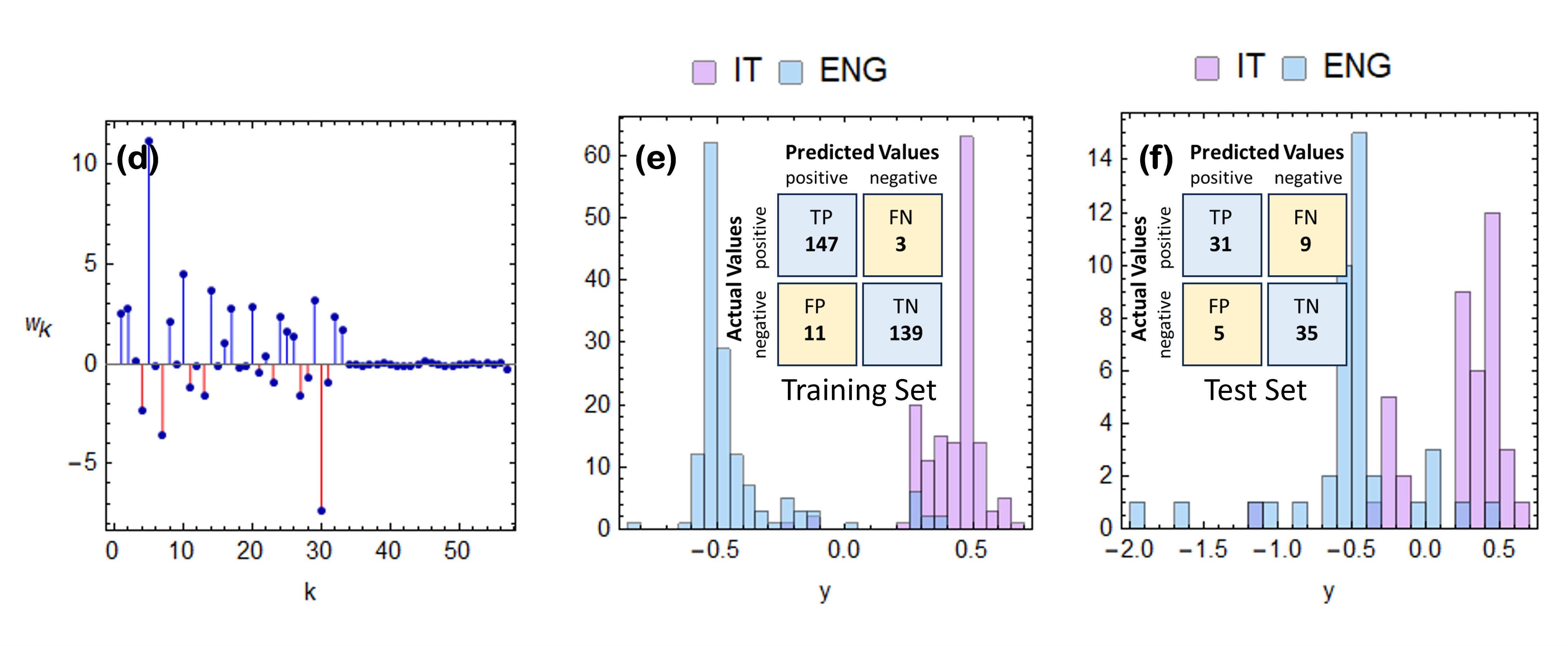}
\caption{Performance of the EQCM architecture on Task~2 (Italian vs English words) with dynamical attention active and identical hyperparameters as in Fig.~\ref{fig2} ($\sigma = 0.1$, $\tau = 10$, $\lambda = 2\times10^{-3}$, $g_1 = 0.1$, $g_2 = 0.4$).\\ 
Panels (a)–(c) correspond to the consonant–vowel encoding already employed in Task~1. Panel (a) shows the learned weights $w_k$, which concentrate on a restricted subset of internal categories. The corresponding training-set performance, shown in panel (b), yields $\mathrm{Accuracy} = 0.9600$ and $\mathrm{BA} = 0.9600$, with $\mathrm{Precision}_{\mathrm{IT}} = 0.9259$ and $\mathrm{Precision}_{\mathrm{ENG}} = 1.000$. On the test set (panel (c)), the model achieves $\mathrm{Accuracy} = 0.9625$ and $\mathrm{BA} = 0.9625$, confirming stable generalization across languages.\\
Panels (d)–(f) report the results obtained using a two-bucket maximum-entropy encoding. In this case, the buckets are constructed in a label-aware fashion by estimating letter frequencies from the Italian portion of the training set and partitioning the alphabet accordingly. Panel (d) displays the corresponding learned weights, while panels (e) and (f) show the training and test performance, respectively. The training set yields $\mathrm{Accuracy} = 0.9533$ and $\mathrm{BA} = 0.9533$, with $\mathrm{Precision}_{\mathrm{IT}} = 0.9308$ and $\mathrm{Precision}_{\mathrm{ENG}} = 0.9789$. On the test set, the performance decreases to $\mathrm{Accuracy} = 0.8250$ and $\mathrm{BA} = 0.8250$, reflecting increased overlap between the class-conditional distributions of the deliberative index $y$.\\
These results indicate that while both encodings allow the architecture to discriminate between Italian and English words with high training accuracy, the consonant–vowel partition provides superior robustness and generalization for this task. The label-aware maximum-entropy buckets, being constructed from empirical letter frequencies estimated on a finite Italian training subset of 150 words, inevitably reflect a sample-dependent approximation of the true dictionary-level statistics. As a consequence, the resulting frequent-symbol bucket may encode a slightly distorted version of the underlying Italian distribution, which reduces its transferability when confronted with a competing structured language such as English.}
\label{fig3}
\end{figure}

Having defined the data set and the two encoding strategies, we now turn to the analysis of the performance of the deliberative system, as summarized in Fig.~\ref{fig3}.

Our discussion starts with the performance analysis, reported in panels (a)-(c), of the consonant-vowel encoding already introduced in Task~1. 
Panel (a), in particular, shows the learned weights $w_k$ with dynamical attention active. 
As in the previous task, the weights concentrate on a restricted subset of internal categories, indicating that the model selects a structured and parsimonious combination of observables.

The training-set performance, summarized in panel (b), is high and well balanced across classes, with 
$\mathrm{Accuracy} = 0.9600$ and $\mathrm{BA} = 0.9600$. 
The class-conditional precisions are $\mathrm{Precision}_{\mathrm{IT}} = 0.9259$ and $\mathrm{Precision}_{\mathrm{ENG}} = 1.0000$, indicating that English words are never misclassified as Italian on the training set, while a small fraction of Italian words is assigned to the English class.

On the test set (panel (c)), the performance remains essentially unchanged, with 
$\mathrm{Accuracy} = 0.9625$ and $\mathrm{BA} = 0.9625$, confirming stable generalization across languages. 
The overlap between the class-conditional distributions of the deliberative index $y$ remains limited, and the confusion matrix shows only one false negative and two false positives. 
Overall, the consonant-vowel encoding provides a robust and transferable representation capable of capturing discriminative regularities shared across distinct structured languages.

Panels (d)-(f) report the results obtained with the data-driven two-bucket maximum-entropy encoding calibrated on the Italian training subset. 
In particular, panel (d) displays the distribution of the learned weights with the dynamical attention active. Despite a qualitative similarity with the distribution shown in panel (a), the weights associated with two internal categories are markedly amplified, indicating their increased relevance in building the deliberative index $y$. A direct comparison between panels (a) and (d), therefore, highlights the role of the encoding strategy in modulating the relative importance of internal categories in task learning.

On the training set (panel (e)), the performance remains high, with 
$\mathrm{Accuracy} = 0.9533$ and $\mathrm{BA} = 0.9533$. 
The class-conditional precisions are $\mathrm{Precision}_{\mathrm{IT}} = 0.9308$ and $\mathrm{Precision}_{\mathrm{ENG}} = 0.9789$, indicating that the model successfully adapts to the empirical statistics encoded in the buckets.

However, on the test set (panel (f)), the performance decreases to 
$\mathrm{Accuracy} = 0.8250$ and $\mathrm{BA} = 0.8250$. 
The increased overlap between the class-conditional distributions of $y$ is reflected in a larger number of both false positives and false negatives, signaling a reduction of the discriminative margin and weaker generalization.

The degradation observed for the maximum-entropy encoding can be traced back to the deliberate choice of constructing the buckets strictly from finite-sample empirical frequencies, without incorporating linguistic corrections for accented vowels not present in the Italian training subset. 
While such letters could be grouped with their unaccented counterparts on phonological grounds, their exclusion induces a sample-dependent distortion of the vowel structure encoded in the frequent bucket. 

This distortion acts as an effective source of encoding noise. 
On the training set, the regression stage compensates for this imperfection, preserving a high classification accuracy. 
On unseen data, however, the misalignment between the learned partition and the underlying phonological regularities reduces the stability of the feature embedding and primarily affects generalization performance.

These results highlight two complementary aspects of the proposed architecture. 
First, the EQCM framework is capable of discriminating between Italian and English words with high accuracy under both encoding strategies. 
Second, encodings that better reflect structurally meaningful and language-agnostic phonological constraints yield superior robustness and transferability, whereas purely data-driven partitions inferred from limited samples may introduce noise that selectively impacts generalization.

\section{Hardware-compatible quantum deliberation systems}
\label{sec:HF}

In order to assess the concrete implementability of the quantum deliberation paradigm on current quantum hardware, we reformulate the internal dynamics in a form compatible with the connectivity constraints of near-term devices. Present-day NISQ architectures typically allow only local two-qubit interactions dictated by the physical layout of the qubits (linear chains or low-dimensional lattices). It is therefore natural to restrict the Hamiltonian to one- and two-body terms acting on nearest neighbours.

With these motivations, we consider a chain of $N=7$ qubits and, consistently with Task~2, the binary classification of seven-letter Italian and English words using the consonant--vowel encoding introduced above. For an input vector $\boldsymbol{z}=(z_1,\dots,z_N)$ produced by such encoding, the total Hamiltonian is written as
\begin{equation}
H(\boldsymbol{z};g_1,g_2)=H_0+H_{I}(\boldsymbol{z};g_1,g_2),
\end{equation}
where the free part is given by
\begin{equation}
H_0=
J \sum_{i=1}^{N-1} \sigma_z^{(i)}\sigma_z^{(i+1)}
+ B_z \sum_{i=1}^{N} \sigma_z^{(i)}
+ B_x \sum_{i=1}^{N} \sigma_x^{(i)},
\end{equation}
and the attention contribution reads
\begin{equation}
H_{I}(\boldsymbol{z};g_1,g_2)
=
- g_1 \sum_{i=1}^{N} z_i\, \sigma_z^{(i)}
- g_2 \sum_{i=1}^{N-1} z_i z_{i+1}\, \sigma_z^{(i)}\sigma_z^{(i+1)}.
\end{equation}

The free Hamiltonian $H_0$ corresponds to a transverse- and longitudinal-field Ising chain, which can be tuned in a non-integrable regime by appropriately selecting the parameters $J$, $B_{x}$ and $B_z$. The nearest-neighbour couplings $\sigma_z^{(i)}\sigma_z^{(i+1)}$ and the single-qubit fields $\sigma_z^{(i)}$ and $\sigma_x^{(i)}$ are directly compatible with hardware supporting local interactions and arbitrary single-qubit rotations. The choice of a non-integrable regime is essential in order to generate non-trivial correlations and entanglement during the exploratory phase, thereby inducing a highly nonlinear feature map from the classical input to the final quantum state. Once the validity of this assumption is guaranteed, no long-range couplings or multi-qubit interactions beyond two-body nearest-neighbour terms are required.

The attention term is constructed according to the same locality principle. The first contribution modulates the local longitudinal fields proportionally to the encoded input components $z_i$, while the second modulates the nearest-neighbour couplings through products $z_i z_{i+1}$. Both terms preserve strict locality and can therefore be implemented without introducing non-native connectivity overhead. In this formulation, the entire Hamiltonian consists exclusively of one- and two-body operators between adjacent qubits, making it directly compatible with present-day architectures.

After unitary evolution for a time $\tau$, the state $\rho(\boldsymbol{z};\tau)$ is mapped to a classical feature vector through expectation values of local and nearest-neighbour observables. Specifically, the feature vector includes the single-site averages $\langle \sigma_x^{(k)} \rangle$, $\langle \sigma_y^{(k)} \rangle$, and $\langle \sigma_z^{(k)} \rangle$ for $k=1,\dots,7$, together with the nearest-neighbour correlators $\langle \sigma_x^{(k)}\sigma_x^{(k+1)} \rangle$, $\langle \sigma_y^{(k)}\sigma_y^{(k+1)} \rangle$, and $\langle \sigma_z^{(k)}\sigma_z^{(k+1)} \rangle$ for $k=1,\dots,6$. Moreover, the expectation value of the identity operator on the full Hilbert space, i.e. $\langle\mathcal{I}\rangle$, is also included as a last component of the feature vector. The readout stage therefore requires only local measurements and two-qubit correlators between adjacent qubits, which are accessible through standard basis rotations and repeated sampling. Moreover, learning remains purely linear in this feature space.

\begin{figure}
\includegraphics[width=1\textwidth]{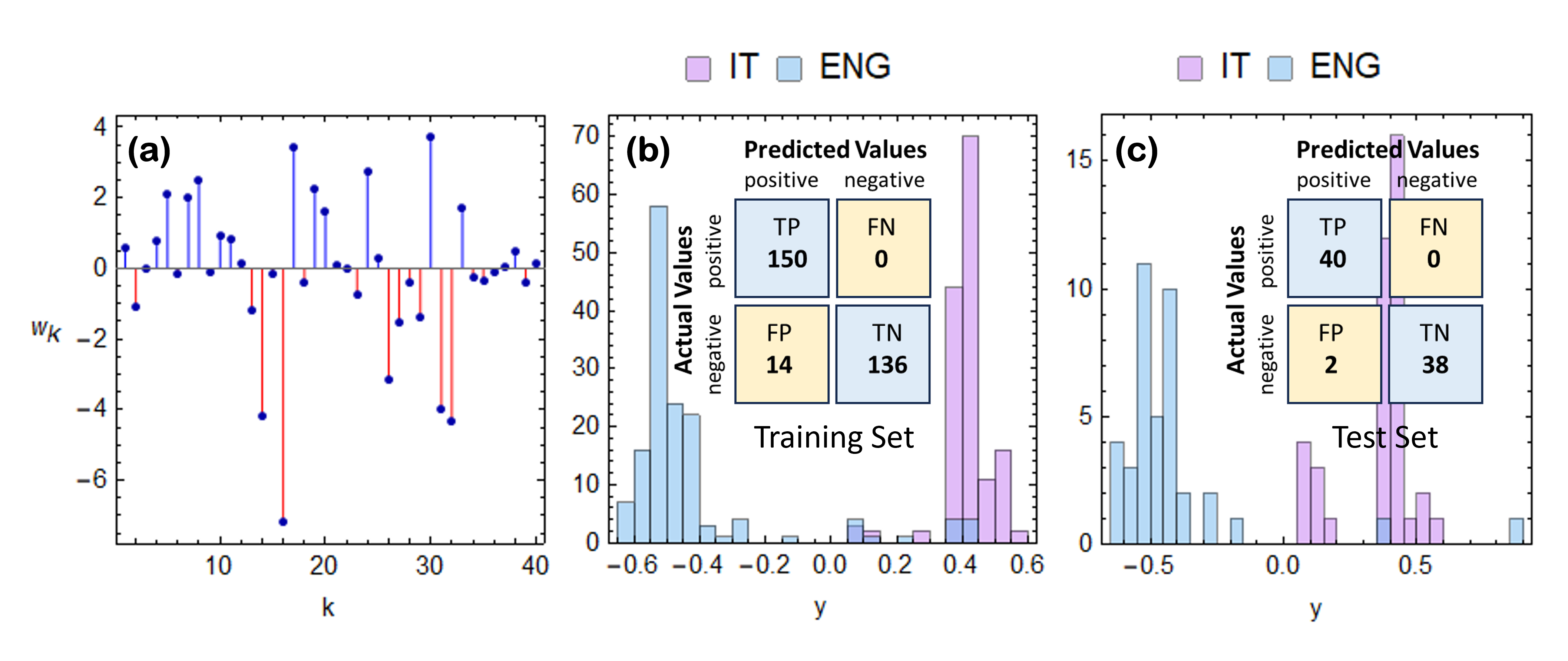}\\
\includegraphics[width=1\textwidth]{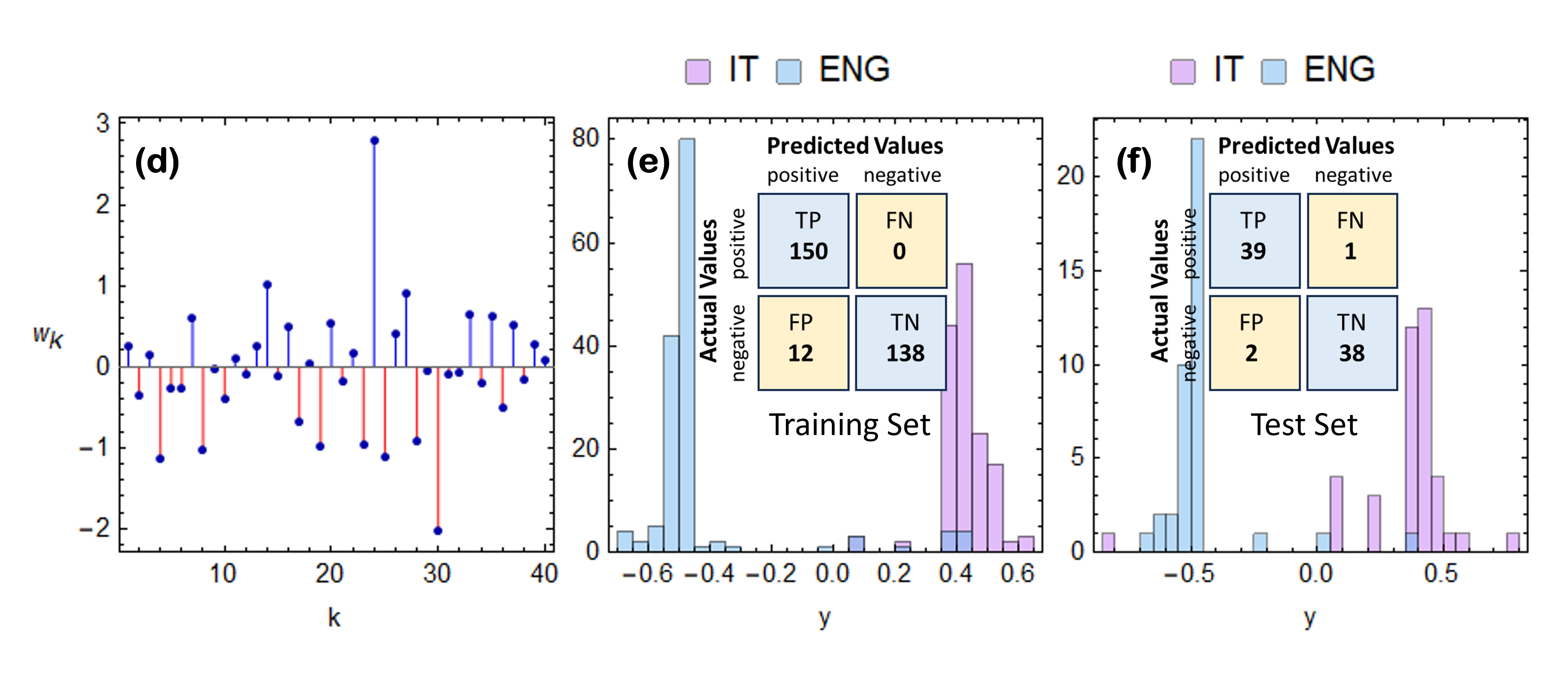}
\caption{Hardware-compatible implementation of the deliberative architecture for Task~2, consisting in the classification of seven-letter Italian and English words using the consonant--vowel encoding. For all the panels we used the following set of parameters: $J=-1$, $B_x=0.7$, $B_z=1.5$, $\tau=20$ and $\lambda=2 \cdot 10^{-3}$. First row (a-c): performance with attention active, $g_1 = g_2 = 2$. Second row (d-f): performance with attention switched off, $g_1 = g_2 = 0$.\\ 
Panels (a,d) report the learned readout weights $w_k$ for the two configurations, 
showing comparable sparsity patterns and magnitude distributions, consistent with an analogous linear recombination of quantum features in both regimes. 
Panels (b,e) display the distributions of the deliberative index $y$ for the training set together with the corresponding confusion matrices.\\ 
With attention active (b), the model achieves $\mathrm{TP}=150$, $\mathrm{TN}=136$, $\mathrm{FP}=14$, $\mathrm{FN}=0$, corresponding to an accuracy of $95.3\%$, precision of $91.5\%$, recall (sensitivity) of $100\%$, and specificity of $90.7\%$.\\ 
In the absence of attention (e), we obtain $\mathrm{TP}=150$, $\mathrm{TN}=138$, $\mathrm{FP}=12$, $\mathrm{FN}=0$, yielding an accuracy of $96.0\%$, precision of $92.6\%$, recall of $100\%$, and specificity of $92.0\%$.\\ 
Panels (c,f) show the corresponding test-set results. 
With attention active (c), the confusion matrix gives $\mathrm{TP}=40$, $\mathrm{TN}=38$, $\mathrm{FP}=2$, $\mathrm{FN}=0$, corresponding to an accuracy of $97.5\%$, precision of $95.2\%$, recall of $100\%$, and specificity of $95.0\%$.\\ 
With attention switched off (f), we find $\mathrm{TP}=39$, $\mathrm{TN}=38$, $\mathrm{FP}=2$, $\mathrm{FN}=1$, corresponding to an accuracy of $96.3\%$, precision of $95.1\%$, recall of $97.5\%$, and specificity of $95.0\%$.\\ 
Overall, the two configurations exhibit essentially overlapping performance on both training and test sets, indicating that, within this hardware-compatible implementation, the inclusion of the dynamical attention term does not produce a measurable advantage in terms of classification accuracy, precision, recall, or specificity.}
\label{fig4}
\end{figure}

Figure~\ref{fig4} reports the performance of this hardware-compatible implementation on Task~2. All simulations are performed by fixing the parameters in $H_0$ to the values used in Ref.~\cite{Geller2022,Xiong2025,DeLorenzis2025}, namely $J=-1$ for the nearest-neighbour Ising coupling, $B_z=1.5$ for the longitudinal field, and $B_x=0.7$ for the transverse field, thereby ensuring a non-integrable dynamical regime, while the evolution time is kept fixed at $\tau=20$.\\ The first row (panels a--c) corresponds to the case with attention active, $g_1=g_2=2$, while the second row (panels d--f) corresponds to the case with attention switched off, $g_1=g_2=0$. Panels (a) and (d) show the learned linear readout weights $w_k$. In both configurations the weight profiles display comparable magnitudes and qualitative structure, indicating that the classifier relies on a similar linear recombination of quantum features irrespective of the presence of the dynamical attention mechanism.

Panels (b) and (e) display the deliberative index distributions on the training set together with the corresponding confusion matrices. With attention active, the model yields $\mathrm{TP}=150$, $\mathrm{TN}=136$, $\mathrm{FP}=14$, and $\mathrm{FN}=0$, corresponding to an accuracy of $95.3\%$, precision of $91.5\%$, recall of $100\%$, and specificity of $90.7\%$. Without attention, one obtains $\mathrm{TP}=150$, $\mathrm{TN}=138$, $\mathrm{FP}=12$, and $\mathrm{FN}=0$, corresponding to an accuracy of $96.0\%$, precision of $92.6\%$, recall of $100\%$, and specificity of $92.0\%$.

Panels (c) and (f) report the performance on the test set. With attention active, the confusion matrix gives $\mathrm{TP}=40$, $\mathrm{TN}=38$, $\mathrm{FP}=2$, and $\mathrm{FN}=0$, corresponding to an accuracy of $97.5\%$, precision of $95.2\%$, recall of $100\%$, and specificity of $95.0\%$. Without attention, one finds $\mathrm{TP}=39$, $\mathrm{TN}=38$, $\mathrm{FP}=2$, and $\mathrm{FN}=1$, corresponding to an accuracy of $96.3\%$, precision of $95.1\%$, recall of $97.5\%$, and specificity of $95.0\%$. The two configurations therefore exhibit essentially overlapping performance on both training and test sets.

These results demonstrate that the quantum deliberation paradigm can be formulated entirely in terms of strictly local Ising-type interactions and local observables, while retaining high classification accuracy and robust generalization. For the present task, the non-integrable exploratory dynamics generated by \(H_0\) is already sufficient to produce a discriminative feature map, and the explicit attention modulation does not lead to a significant improvement. This finding does not contradict the results obtained with a fully nonlocal random \(H_0\), where dynamical attention plays a crucial role in enhancing classification performance. Rather, it indicates that the relevance of attention depends on the structure of the underlying Hamiltonian. Indeed, it provides an important inductive bias for highly nonlocal dynamics, while its contribution can be less pronounced when the intrinsic dynamics is spatially local. The latter setting is particularly relevant for present-day quantum computers, whereas architectures supporting more general nonlocal interactions may become increasingly accessible with future quantum technologies.

\section{Spectral characterization of the deliberative operator}
\label{sec:DelOp}

So far, the analysis has focused on the deliberative index $y$ and on the capability of the EQCM architecture to reproduce the labels associated with different classes. In particular, we have shown that the learning procedure determines a set of weights $w_k$ such that the expectation value of the learned operator $O_{\boldsymbol{w}}$ yields a continuous index whose distribution correlates with the target labels, thereby enabling accurate classification.

From the perspective of quantum cognition, the learned operator $O_{\boldsymbol{w}}$ can be viewed as a newly constructed conceptual category, obtained by combining pre-existing internal categories $Q_k$. In biological cognitive systems, such internal categories are shaped by accumulated experience and cognitive organization, providing the prior conceptual structure upon which new deliberative processes are built. The learning stage therefore produces a new deliberative observable that synthesizes these internal categories into an effective evaluative operator adapted to the task.

Once the operator $O_{\boldsymbol{w}}$ has been learned, the presentation of an input induces a quantum evolution of the initial state, leading at time $\tau$ to a mental state described by $\rho(z;\tau)$.  In a single-shot measurement, the mental state would probabilistically collapse onto one of the eigenstates of the deliberative operator, with probability determined by its projection onto the corresponding eigenspace. From this point of view, the deliberative process can be interpreted as a probabilistic selection among the elementary evaluative outcomes associated with the eigenvalues of $O_{\boldsymbol{w}}$.

This interpretation suggests that inputs belonging to the same class may preferentially populate specific regions of the Hilbert space, corresponding to subsets of eigenstates of $O_{\boldsymbol{w}}$. In other words, one may expect the emergence of effective \emph{deliberative subspaces}, which are more strongly activated by inputs of a given class. The experimentally accessible quantity $y$ then arises as the expectation value of $O_{\boldsymbol{w}}$, i.e. as the average outcome of repeated measurements performed under identical conditions.

From this perspective, it becomes natural to investigate whether, and to what extent, the learned operator $O_{\boldsymbol{w}}$ induces a structured decomposition of the Hilbert space into such deliberative subspaces. This issue motivates a systematic spectral analysis of $O_{\boldsymbol{w}}$, aimed at identifying the general properties of its eigenvalues and eigenvectors, and clarifying the mechanisms through which the class-dependent features are encoded in the quantum state.\\
In the following, we formalize this analysis by deriving general results that are independent of the specific task and provide a principled interpretation of the deliberative index.

To begin the spectral analysis, let us recall that the learned deliberative operator is defined as
\begin{equation}
O_{\boldsymbol{w}} = \sum_k w_k Q_k,
\end{equation}
where the coefficients $w_k \in \mathbb{R}$ are determined by the learning procedure, while the operators $Q_k$ belong to the fixed set of internal categories introduced in the previous sections. In the present framework, each $Q_k$ is taken to be, in general, a Pauli string, namely a tensor product of single-qubit operators chosen from the set $\{I,\sigma_x,\sigma_y,\sigma_z\}$. Since  $O_{\boldsymbol{w}}$ is Hermitian by construction, it admits a spectral decomposition of the form
\begin{equation}
O_{\boldsymbol{w}} = \sum_{i=1}^{D} \lambda_i \, |\psi_i\rangle\langle \psi_i| ,
\end{equation}
where the operator $O_{\boldsymbol{w}}$ has been expressed in terms of its real eigenvalues $\lambda_i$ and of its eigenvectors ${|\psi_i\rangle}$, which form an orthonormal basis of the $D$-dimensional Hilbert space. The spectral decomposition allows one to express the deliberative index
\begin{equation}
y = \mathrm{Tr}\!\left[\rho(z;\tau)\, O_{\boldsymbol{w}}\right]
\end{equation}
in terms of eigenvalues and eigenvectors of the learned deliberative operator. Indeed, using  the spectral representation of $O_{\boldsymbol{w}}$, one easily obtains:
\begin{equation}
y = \sum_{i=1}^{D} \lambda_i \, p_i ,
\end{equation}
where $p_i = \langle \psi_i | \rho(z;\tau) | \psi_i \rangle $. The quantities $p_i$ satisfy
\begin{equation}
p_i \ge 0,
\qquad
\sum_{i=1}^{D} p_i = 1,
\end{equation}
and therefore define a probability distribution over the eigenstates of $O_{\boldsymbol{w}}$. From the standpoint of quantum cognition, $p_i$ represents the probability that, in a single projective measurement of the deliberative operator, the mental state $\rho(z;\tau)$ collapses onto the eigenstate $|\psi_i\rangle$, yielding the corresponding elementary deliberative outcome $\lambda_i$.

Since $y$ represents the expectation value associated with an ensemble of projective measurements of the deliberative observable, the index $y$ is a convex combination of the eigenvalues of the deliberative operator. Thus, denoting by $\lambda_{\min}$ and $\lambda_{\max}$ the minimum and maximum eigenvalues of $O_{\boldsymbol{w}}$, respectively, one necessarily has
\begin{equation}
\lambda_{\min} \le y \le \lambda_{\max}.
\end{equation}
Thus, the spectrum of the deliberative operator fixes the maximal range accessible to the deliberative index, while the actual value of $y$ depends on the cognitive weight $p_i$ assigned by the evolved mental state to the elementary deliberative outcomes $\lambda_i$.

A further quantity of interest is the mean value $\overline{\lambda}$ of the spectrum of the deliberative operator, defined as the average over all its eigenvalues, namely
\begin{equation}
\overline{\lambda}=\frac{1}{D} \sum_{i=1}^{D} \lambda_i = \frac{\mathrm{Tr}(O_{\boldsymbol{w}})}{D}.
\end{equation}
The quantity $\bar{\lambda}$ depends on the trace of $O_{\boldsymbol{w}}$, which can be evaluated explicitly by exploiting the tensor-product structure of the operators $Q_k$. Using the linearity of the trace, one has
\begin{equation}
\mathrm{Tr}(O_{\boldsymbol{w}}) = \sum_k w_k \, \mathrm{Tr}(Q_k).
\end{equation}
Moreover, for Pauli strings $\tau_1 \otimes \cdots \otimes \tau_N$, the trace factorizes as
\begin{equation}
\mathrm{Tr}\!\left( \tau_1 \otimes \cdots \otimes \tau_N \right)
=
\prod_{j=1}^{N} \mathrm{Tr}(\tau_j),
\end{equation}
where $\tau_j \in \{I,\sigma_x,\sigma_y,\sigma_z\}$. Since $\mathrm{Tr}(\sigma_x) = \mathrm{Tr}(\sigma_y) = \mathrm{Tr}(\sigma_z) = 0$ and $
\mathrm{Tr}(I) = 2$, every Pauli string has vanishing trace unless all its factors are equal to the identity, so that we obtain:
\begin{equation}
\mathrm{Tr}(Q_k) = 0
\quad \text{for all } Q_k \neq I^{\otimes N},
\end{equation}
while $\mathrm{Tr}(I^{\otimes N}) = 2^N = D$. As a consequence, if the identity string $I^{\otimes N}$ is included among the operators $Q_k$, with associated coefficient $w_M$, one obtains $\mathrm{Tr}(O_{\boldsymbol{w}}) = w_M \, D$ and therefore $\overline{\lambda}=w_M$. In the opposite case,  $\mathrm{Tr}(O_{\boldsymbol{w}})=0$ and the spectrum has zero mean (i.e. $\overline{\lambda}=0$). Thus, we have demonstrated that the average value of the spectrum of the deliberative operator is entirely controlled by the coefficient of the identity operator, while the presence of all other Pauli strings only affects the eigenvalues dispersion around their mean.

The previous result also has a direct implication for the qualitative structure of the spectrum. Unless all eigenvalues vanish identically, which is not expected for a meaningful deliberative operator, the condition $\mathrm{Tr}(O_{\boldsymbol{w}})=0$ necessarily implies the coexistence of both positive and negative eigenvalues, giving rise to distinct spectral sectors.

Moreover, even when $\mathrm{Tr}(O_{\boldsymbol{w}}) \neq 0$, the presence of  eigenvalues of opposite sign remains a generic feature of a nontrivial deliberative operator, under the requirement that the index $y$ reproduces target values of opposite sign for different classes.

Despite the above arguments,  the distribution of the eigenvalues of the deliberative operator $O_{\boldsymbol{w}}$ is determined by the learned weights $w_k$ and, in general, does not exhibit any symmetry with respect to zero. Even in the case $\mathrm{Tr}(O_{\boldsymbol{w}})=0$, which ensures that the spectrum has zero mean, no pairwise symmetry $\lambda \leftrightarrow -\lambda$ is implied.

An exact symmetry of the spectrum with respect to zero requires a stronger structural condition. In particular, suppose that there exists a unitary and Hermitian operator $\Gamma$ such that
\begin{equation}
\Gamma^2 = I,
\end{equation}
\begin{equation}
\Gamma O_{\boldsymbol{w}} \Gamma = - O_{\boldsymbol{w}}.
\end{equation}
Then the spectrum of $O_{\boldsymbol{w}}$ is symmetric, which implies that if $\lambda$ is an eigenvalue, then $-\lambda$ is also an eigenvalue with the same multiplicity.

Indeed, if
\begin{equation}
O_{\boldsymbol{w}} |\psi\rangle = \lambda |\psi\rangle,
\end{equation}
then
\begin{equation}
O_{\boldsymbol{w}} (\Gamma |\psi\rangle) = -\Gamma O_{\boldsymbol{w}} |\psi\rangle = -\lambda (\Gamma |\psi\rangle),
\end{equation}
showing that $\Gamma|\psi\rangle$ is an eigenvector with eigenvalue $-\lambda$.

This condition is analogous to a chiral symmetry and is not generically satisfied by a linear combination of Pauli strings with arbitrary coefficients. Therefore, in the absence of additional constraints on the set $\{Q_k\}$ and on the weights $w_k$, one should not expect the spectrum of $O_{\boldsymbol{w}}$ to be exactly symmetric, even when its mean vanishes.

One could, in principle, enforce such a symmetry by restricting the admissible mental categories $\{Q_k\}$ so that the learned deliberative operator $O_{\boldsymbol{w}}$ anticommutes with a symmetry operator $\Gamma$. For example, choosing
\begin{equation}
\Gamma=\sigma_x\otimes I^{\otimes (N-1)},
\end{equation}
the condition
\begin{equation}
\{\Gamma,O_{\boldsymbol{w}}\}=0
\end{equation}
implies that only Pauli strings whose first factor is $\sigma_y$ or $\sigma_z$ can contribute to the deliberative operator, since these operators change sign under conjugation by $\sigma_x$. The resulting operator therefore takes the form
\begin{equation}
O_{\boldsymbol{w}}=
\sum_{\alpha} w_\alpha^{(z)}(\sigma_z\otimes P_\alpha)
+
\sum_{\beta} w_\beta^{(y)}(\sigma_y\otimes P_\beta),
\end{equation}
where $P_\alpha$ and $P_\beta$ are arbitrary Pauli strings acting on the remaining qubits. In this way, the symmetry constraint selects a restricted family of mental categories and guarantees an exactly symmetric deliberative spectrum.

However, it is not \emph{a priori} clear whether imposing such structural constraints preserves the expressive power of the model, since it reduces the accessible operator subspace. In particular, enforcing an exactly symmetric spectrum may limit the capability of the learned operator to represent intrinsically asymmetric decision landscapes.\\
On the other hand, symmetry-constrained constructions could provide a natural framework for the development of physics-informed deliberative architectures \cite{Karniadakis2021}.

We now turn to the estimation of the spectral radius of the deliberative operator. To this end, it is convenient to introduce the operator norm and some of its basic properties.

For a linear operator $A$ acting on a Hilbert space, the operator norm is defined as
\begin{equation}
\|A\| = \sup_{\|\phi\|=1} \|A\phi\|,
\end{equation}
and  satisfies the standard norm properties:
\begin{equation}
\|cA\| = |c|\,\|A\|,
\end{equation}
\begin{equation}
\|A+B\| \le \|A\| + \|B\|,
\end{equation}
\begin{equation}
\|AB\| \le \|A\|\,\|B\|.
\end{equation}
Moreover, the norm of the operator $A\otimes B$ is given by:
\begin{equation}
\|A\otimes B\| = \|A\|\,\|B\|.
\end{equation}
If $A$ is Hermitian, the operator norm coincides with the largest absolute value of its eigenvalues,
\begin{equation}
\|A\| = \max_i |\mu_i|,
\end{equation}
where $\{\mu_i\}$ denotes the spectrum of $A$.\\ 
Using the operator norm properties, an estimate of the spectral radius of  $O_{\boldsymbol{w}}$ can be obtained.  To this purpose, let us consider a generic Pauli string $Q_k$. Since every Pauli matrix squares to the identity, one has $(Q_k)^2=I^{\otimes N}$, which implies that the eigenvalues of $Q_k$ are restricted to $q=\pm1$.  Therefore, the spectrum of any Pauli string is contained in the set $\{+1,-1\}$, and its operator norm is $\|Q_k\| = 1$. Thus, using the triangle inequality and the fact that each Pauli string has unit norm, one finds
\begin{equation}
\|O_{\boldsymbol{w}}\|
=
\left\|\sum_k w_k Q_k\right\|
\le
\sum_k |w_k|\,\|Q_k\|
=
\sum_k |w_k|.
\end{equation}
Since $O_{\boldsymbol{w}}$ is Hermitian, this implies the bound
\begin{equation}
\max_i |\lambda_i|
\le
\sum_k |w_k|,
\end{equation}
where $\{\lambda_i\}$ are the eigenvalues of $O_{\boldsymbol{w}}$. Hence, the spectral radius of the deliberative operator is controlled by the $\ell^1$ norm of the learned weight vector and shows the tendency to increase with the cardinality of the operator basis retained in the readout layer. Of course, the bound given above is generally not saturated, since cancellations due to the noncommutativity of the different Pauli strings may substantially reduce the actual spectral radius. Nevertheless, we conclude that increasing the expressive complexity of the readout enlarges the range of possible eigenvalues of the learned deliberative operator.

The obtained spectral radius estimate immediately implies that an arbitrary eigenvalue $\lambda_i$ of the deliberative operator is subject to the following inequality:
\begin{equation}
|\lambda_i|\le \|O_{\boldsymbol{w}}\| \le \sum_k |w_k|.
\end{equation}
A further spectral characterization can be obtained by considering the variance $\sigma_{\lambda}^2$ of the eigenvalue distribution. Exploiting the Hilbert-Schmidt orthogonality of Pauli strings,
\begin{equation}
\mathrm{Tr}(Q_k Q_{k'}) = D\,\delta_{kk'},
\end{equation}
one obtains:
\begin{equation}
\overline{\lambda^2}
=
\frac{1}{D}\sum_i \lambda_i^2
=
\frac{\mathrm{Tr}(O_{\boldsymbol{w}}^2)}{D}
=
\sum_k w_k^2 ,
\end{equation}
where $D$ is the Hilbert-space dimension. As a consequence, the variance of the eigenvalue distribution $\sigma_\lambda^2$ is directly determined by the weight vector,
\begin{equation}
\sigma_\lambda^2
=
\overline{\lambda^2}-( \ \overline{\lambda} \ )^2
=
\sum_{k \neq M} w_k^2,
\end{equation}
while the corresponding root-mean-square spectral scale is given by
\begin{equation}
\sigma_\lambda
=
\sqrt{\sum_{k \neq M} w_k^2} \le \sum_{k \neq M}|w_k|.
\end{equation}
This observation provides an additional interpretation of the role played by Tikhonov regularization within the ridge-regression learning procedure. Indeed, the regularization term penalizes large values of the weight vector norm,
\begin{equation}
\sum_k w_k^2,
\end{equation}
which, according to the previous relations, directly controls the variance of the deliberative spectrum. As a consequence, ridge regression naturally favors learned deliberative operators characterized by a reduced spectral width, thereby suppressing excessively large fluctuations of the deliberative index. From this perspective, the regularization procedure can be interpreted as driving the model toward an optimal-variance representation, balancing expressive capability and spectral stability.

Up to this point, the analysis has focused on spectral properties of the deliberative operator that can be characterized in a model-independent way, such as the spectral mean value, variance, and support. A more detailed description of the spectrum would, in general, require explicit knowledge of the specific mental categories entering the construction of the model. Nevertheless, further general considerations can still be formulated in the opposite limit in which the set of admissible mental categories becomes close to maximal, densely spanning the space of Pauli strings.

When the set of admissible mental categories becomes sufficiently large to densely sample the space of Pauli strings, one may conjecture that the learned deliberative operator approaches the behavior of a generic random Hermitian matrix. In such a regime, the spectral properties of $O_{\boldsymbol{w}}$ are expected to become increasingly universal, with the eigenvalue distribution plausibly converging toward a Wigner semicircle law \cite{Tao2012}, largely independently of the specific task. Under this perspective, excessive proliferation of mental categories could progressively wash out task-dependent spectral structures, driving the deliberative operator toward a universal random-matrix regime.

In practice, however, the performance of extreme-learning architectures does not typically collapse \cite{Fujii2017,Settino2024,jing2025} as the number of available features increases, but rather tends to saturate beyond a certain threshold. Within the present framework, this behavior may admit a natural spectral interpretation associated with the ridge-regression stage. Indeed, the regularization procedure suppresses non-informative or redundant mental categories by driving many coefficients toward negligible values, thereby selecting an effective low-dimensional subspace of relevant observables. As a consequence, the learned deliberative operator avoids entering a fully universal random-matrix regime and retains the task-dependent spectral structures necessary for discrimination.

From a dynamical perspective, this mechanism may also be interpreted as an effective reduction of the accessible informational sector. Although the global quantum evolution remains unitary, restricting the readout to a reduced subset of relevant categories effectively resembles the monitoring of a subsystem, where part of the available information is discarded. Such a restriction induces a form of effective dissipative coarse graining in the accessible information dynamics. Since weak dissipation and partial information loss are known to play a beneficial role in reservoir and extreme-learning architectures \cite{Sannia2024}, ridge regression may effectively stabilize the learning dynamics by balancing expressive complexity against excessive spectral universality. Within this perspective, the empirically observed saturation of performance with increasing feature-space dimension emerges as a natural consequence of the regularized spectral selection performed by the learning stage.

 We now turn to the distribution of the spectral weights $p_i = \langle \psi_i | \rho(z;\tau) | \psi_i \rangle$, which encode how the evolved state populates the eigenbasis of the deliberative operator. In general, these probabilities are not directly constrained by the algebraic structure of $O_{\boldsymbol{w}}$, but are instead determined by the interplay between the input-dependent dynamics and the optimization of the readout layer.

It is important to observe that the loss function minimized in the ridge regression step imposes only a weak constraint on the spectral structure of $O_{\boldsymbol{w}}$. In particular, the optimization targets the expectation value
\begin{equation}
y = \sum_i \lambda_i p_i,
\end{equation}
without explicitly constraining the individual probabilities $p_i$ or enforcing any particular organization of the Hilbert space into sharply defined subspaces. As a consequence, one should not expect, in general, the emergence of strictly separated or well-defined deliberative subspaces associated with disjoint sets of eigenstates. Rather, the model is free to exploit the full spectral richness of $O_{\boldsymbol{w}}$, distributing probability over a large portion of the spectrum in a way that optimally reproduces the target labels.

A further insight can be obtained by considering the relative scale of the target values and of the spectral radius of the deliberative operator. As discussed above, the spectral radius $\|O_{\boldsymbol{w}}\|$ typically grows with the complexity of the model, and can be substantially larger than the target values associated with the different classes. In such a regime, where
\begin{equation}
|y| \ll \|O_{\boldsymbol{w}}\|,
\end{equation}
the deliberative index must arise from a delicate balance between contributions of opposite sign. More precisely, one has
\begin{equation}
y = \sum_i \lambda_i p_i = \sum_{\lambda_i>0} \lambda_i p_i + \sum_{\lambda_i<0} \lambda_i p_i,
\end{equation}
so that the final value of $y$ results from a partial cancellation between positive and negative spectral sectors.

In this situation, eigenstates associated with large values of $|\lambda_i|$, i.e. those located near the spectral edges, acquire a particularly significant role. Indeed, for such states, even small variations in the corresponding probabilities $p_i$ can produce finite changes in the value of $y$. Conversely, eigenstates in the central region of the spectrum, where $|\lambda_i|$ is small, have a weaker leverage on the deliberative index. 

This picture is fully consistent with the absence of sharp spectral constraints imposed by the learning procedure. Indeed, rather than selecting a small number of preferred states, the EQCM architecture exploits a broad superposition of spectral components, with the extremal regions acting as sensitive control parameters for the final outcome of the deliberation.

Remarkably, the above theoretical picture is fully corroborated by the numerical results (not reported here) obtained in the hardware-compatible setting, where both the spectral distribution and the occupation probabilities display the expected structure and confirm the role of extremal eigenvalues in the deliberative process.

\section{Discussion and conclusions}
\label{sec:CONCL}

In this work, we have introduced EQCMs as a class of learning architectures explicitly grounded in the quantum cognition paradigm and implemented within the operational framework of quantum extreme learning. The central idea is to represent inputs as maximum-entropy density matrices $\rho_0(\boldsymbol{z})$ compatible with input-dependent local constraints, to process them through fixed quantum dynamics that generate a nonlinear feature embedding, and to confine learning to a linear combination of expectation values of pre-defined observables (internal categories). In this way, a deliberative index can be defined as the expectation value of a learned deliberative operator $O_{\boldsymbol{w}}$ evaluated on the mental quantum state, described by $\rho(\boldsymbol{z}; \tau)$. The operator $O_{\boldsymbol{w}}$ is expressed in terms of trainable weights $w_k$ inferred from data.

From a conceptual perspective, the architecture realizes a concrete synthesis of three ingredients. First, the maximum-entropy encoding ensures that the initial state represents the least biased quantum description consistent with the available information, in full accordance with Jaynes’ principle \cite{Jaynes57} and with the probabilistic structure of quantum cognition (see Appendix \ref{AppA}). Second, the unitary dynamics generated by $H_0 + H_I$ implements a coherent quantum walk in the internal state space, where exploratory mixing and input-dependent attention shape a high-dimensional nonlinear embedding. Third, the deliberative operator $O_w$ learned by ridge regression encodes the task-specific aggregation of internal categories without altering the underlying dynamical laws. This separation between fixed internal structure and linear task-level adaptation mirrors the distinction, emphasized in quantum cognition, between stable conceptual operators and context-dependent evaluative outcomes.

The numerical experiments on hard deliberation tasks demonstrate that this architecture is capable of extracting relational structure from severely coarse-grained symbolic inputs. In both linguistic benchmarks considered in this work, discrimination does not rely on local symbol identity but on global correlations induced by the encoding and processed by the quantum dynamics. The results show that dynamical attention can act as a physically interpretable inductive bias, selectively enhancing relevant internal categories and improving generalization. At the same time, the architecture remains robust under hardware-constrained implementations based solely on local Ising-type interactions and nearest-neighbour correlators, as discussed in Sec.~\ref{sec:HF}. This establishes that the quantum deliberation paradigm is not tied to fully nonlocal Hamiltonians, but can be reformulated in terms directly compatible with present-day NISQ connectivity constraints.

Beyond the specific benchmarks considered here, the EQCM framework suggests a general methodology for tasks characterized by informational scarcity, label noise, and relational structure. Because the model provides a continuous deliberative index rather than a sharp categorical assignment, it is naturally suited to applications in which uncertainty, ambiguity, and graded evaluation are intrinsic. In medical image analysis, for instance, diagnostic decisions often emerge from subtle correlations among spatially distributed features rather than from isolated local cues. An EQCM-like architecture could extract classical features from different regions of an image, encode them into a maximum-entropy density matrix subject to constraints, process this representation through structured local Hamiltonians, and produce a deliberative index reflecting the aggregated evidence for pathology. Similar considerations apply to environmental monitoring and distributed sensing, where measurements from multiple spatial or spectral channels must be integrated into a robust anomaly score under noisy and partially contradictory conditions.

More broadly, the encoding strategy based on maximum-entropy constraints and the emphasis on relational observables make the framework particularly attractive for sequence analysis in biological contexts. Protein primary sequences, DNA and RNA strings, and other symbolic biological data are naturally amenable to dichotomic or multi-bin encodings reflecting biochemical or structural attributes. The subsequent quantum evolution can generate entangled and correlated features that capture collective patterns beyond local motifs. In this perspective, quantum deliberation becomes a principled mechanism for integrating distributed biochemical evidence into a scalar functional index, potentially relevant for tasks such as functional annotation, mutation impact assessment, or structural class discrimination. Although the present study remains at the level of proof-of-principle simulations, the formal structure of the architecture is fully compatible with scaling strategies based on local Hamiltonians and sparse measurement sets.

An important structural aspect of the EQCM model is its intrinsic scalability in the sense of extreme learning. The quantum subsystem implements a fixed nonlinear map from classical inputs to expectation-value features, and the training stage reduces to a convex linear optimization problem. As a consequence, retraining under new labels or evolving datasets requires no modification of the quantum dynamics and can be performed efficiently. This property is particularly relevant in agent-based or adaptive settings, where rapid updates of deliberative criteria are required while preserving a stable internal conceptual space. Moreover, the proposed EQCM architecture exhibits classification performance comparable to that of an optimized classical Extreme Learning Machine, as demonstrated by the feature-dimension scaling analysis presented in Appendix \ref{AppC}.\\
From a hardware standpoint, the restriction to one- and two-body nearest-neighbor interactions and to local or short-range measurements (possibly implementing the classical shadow strategy \cite{huang2020predicting}) ensures that implementations on superconducting, trapped-ion, or other NISQ platforms are, in principle, feasible within current technological constraints.

It is worth highlighting that the value of the model is not solely based on potential quantum speedups. Even when simulated classically for moderate system sizes, the architecture provides a conceptually coherent and physically interpretable realization of quantum cognition principles within a machine-learning framework. The representation of mental states as density matrices, the treatment of questions as observables, the role of non-commutativity and order effects in focused processing, and the emergence of decisions as expectation values are not merely metaphors but operational components of the algorithm. In this sense, the EQCM constitutes a model of quantum deliberation in its own right, independently of hardware considerations.

Several directions naturally follow from the present work. On the theoretical side, one may analyze the expressive power of the induced feature map in relation to properties of the underlying Hamiltonian, such as integrability versus non-integrability, spectral statistics, and entanglement generation. On the algorithmic side, natural extensions of the present paradigm include incorporating suitable sequences of weak measurements \cite{PhysRevLett.60.1351,RevModPhys.86.307} within the dynamics as an additional computational resource, as well as generalizing the encoding stage to handle raw data with non-dichotomic decision elements through embeddings into appropriate vector spaces, in the spirit of modern language-model representation \cite{Mikolov2013}. On the applied side, systematic benchmarks on structured image datasets, multimodal sensing problems, and biological sequence corpora will be necessary to assess the practical advantages and limitations of the approach.

In conclusion, EQCMs provide a unified framework in which quantum cognition and quantum extreme learning converge into a concrete architecture for deliberative decision making. The model is conceptually grounded, computationally tractable, and compatible with the locality and connectivity constraints of currently available quantum hardware. These features make the proposed framework a promising platform for exploring both quantum-inspired and quantum-native approaches to complex decision tasks, opening a pathway toward scalable deliberative models applicable across a wide range of data-driven scientific problems.

\appendix
\section{Maximum-entropy construction of the initial state}
\label{AppA}

Let $\varrho$ be a density matrix acting on a finite-dimensional Hilbert space $\mathcal{H}$, and let $\{\mathcal{O}_\alpha\}_{\alpha=1}^{M}$ be a set of Hermitian operators representing evaluative observables. Given prescribed expectation values
\[
\mathrm{Tr}(\varrho\,\mathcal{O}_\alpha)=m_\alpha ,
\qquad \alpha=1,\dots,M ,
\]
the maximum-entropy principle in the sense of Jaynes \cite{Jaynes57} selects the quantum state $\varrho^\star$ that maximizes the von Neumann entropy
\[
S(\varrho)=-\mathrm{Tr}(\varrho\ln\varrho),
\]
under the constraints of normalization (i.e., $\mathrm{Tr}(\varrho)=1$) and fixed expectation values.

Introducing Lagrange multipliers $\lambda_0,\lambda_1,\dots,\lambda_M$, the constrained optimization problem can be formulated through the functional
\[
\mathcal{L}[\varrho]
=
-\mathrm{Tr}(\varrho\ln\varrho)
-\lambda_0\bigl(\mathrm{Tr}\,\varrho-1\bigr)
-\sum_{\alpha=1}^{M}\lambda_\alpha
\bigl(\mathrm{Tr}(\varrho\mathcal{O}_\alpha)-m_\alpha\bigr).
\]
Taking the functional derivative of $\mathcal{L}$ with respect to $\varrho$, also considering the relation 
\[\delta \ \mathrm{Tr}(\varrho\ln\varrho)=\mathrm{Tr}((\ln \varrho+\mathcal{I})\delta \varrho), 
\]
and setting it to zero yields the stationary condition
\[
\ln\varrho + (1 + \lambda_0)\mathcal{I}
+\sum_{\alpha=1}^{M}\lambda_\alpha\mathcal{O}_\alpha
=0,
\]
which leads to the formal solution
\[
\varrho^\star
=
\frac{e^{-\sum_{\alpha}\lambda_\alpha\mathcal{O}_\alpha}}{Z},
\qquad
Z=\mathrm{Tr}\!\left(
e^{-\sum_{\alpha}\lambda_\alpha\mathcal{O}_\alpha}
\right).
\]

The Lagrange multipliers $\{\lambda_\alpha\}$ are determined implicitly by enforcing the constraints
\[
\mathrm{Tr}\!\left(
\varrho^\star\mathcal{O}_\alpha
\right)=m_\alpha .
\]
In the general case, when the operators $\mathcal{O}_\alpha$ do not commute, these equations constitute a set of coupled nonlinear equations. As a consequence, the explicit determination of $\varrho^\star$ typically requires numerical methods and does not admit a closed-form solution. This feature reflects the genuinely noncommutative nature of quantum probability and the emergence of correlations induced by constraints written in terms of incompatible observables.

The construction simplifies considerably when the constraint operators commute pairwise,
\[
[\mathcal{O}_\alpha,\mathcal{O}_\beta]=0
\qquad
\forall\,\alpha,\beta .
\]
In this case, all $\mathcal{O}_\alpha$ can be simultaneously diagonalized and the operator, defining the formal solution $\varrho^\star$, factorizes accordingly. This situation corresponds to the assumptions adopted in the main text for the initialization of the mental state.

Specializing to a system of $N$ qubits with Hilbert space
$\mathcal{H}=(C^2)^{\otimes N}$,
we consider local observables
\[
\mathcal{O}_k = \sigma_z^{(k)},
\qquad k=1,\dots,N ,
\]
and impose the constraints
\[
\mathrm{Tr}(\varrho\,\sigma_z^{(k)})=m_k ,
\qquad m_k\in[-1,1].
\]
Under these assumptions, the maximum-entropy solution factorizes as
\[
\varrho^\star
=
\bigotimes_{k=1}^{N}\varrho_k(m_k),
\]
where each single-qubit state takes the form
\[
\varrho_k(m_k)
=
\frac{e^{-\lambda_k\sigma_z}}{\mathrm{Tr}(e^{-\lambda_k\sigma_z})}
=
\frac{1}{2}\bigl(I+m_k\sigma_z\bigr),
\qquad
m_k=-\tanh\lambda_k .
\]

This state is the unique maximum-entropy state compatible with the specified local expectation values. From an information theory perspective, it represents the least structured quantum state consistent with the available data, containing no correlations beyond those explicitly imposed by the constraints. In the context of the present work, this construction provides a natural representation of an unbiased initial mental state encoding the input information solely through its prescribed local averages.

\section{Binary-class confusion matrix and performance metrics}
\label{AppB}

In this Appendix we summarize the classification metrics employed throughout the manuscript for binary discrimination tasks. We adopt the convention that the positive class ($+$) corresponds to label $t_+$ and the negative class ($-$) to label $t_-$. Given a set of $N$ labeled samples, let $y$ denote the continuous deliberative index produced by the model and $\hat t\in\{t_+,t_-\}$ the predicted label obtained through a fixed decision rule. In the present work we adopt the simple sign-based rule $\hat t=t_+$ if $y>0$ and $\hat t=t_-$ if $y<0$.

The performance of the classifier is conveniently summarized by the confusion matrix, which counts how predicted labels compare with the true ones. Four basic integer quantities are defined. True positives (TP) denote the number of samples belonging to the positive class that are correctly predicted as positive. False negatives (FN) correspond to positive samples incorrectly predicted as negative. False positives (FP) denote negative samples that are incorrectly classified as positive, while true negatives (TN) correspond to negative samples correctly identified as such. These four counts exhaust the dataset and therefore satisfy
\[
N=\mathrm{TP}+\mathrm{TN}+\mathrm{FP}+\mathrm{FN}.
\]
It is also convenient to introduce the class supports
\[
N_+ = \mathrm{TP}+\mathrm{FN},\qquad N_-=\mathrm{TN}+\mathrm{FP},
\]
which represent the total number of positive and negative samples in the dataset, respectively.

From these basic quantities, one constructs several standard rates used to characterize the performance of a binary classifier. The true positive rate (TPR), also called recall or sensitivity, measures the fraction of positive samples that are correctly identified:
\[
\mathrm{TPR}=\mathrm{Recall}=\frac{\mathrm{TP}}{\mathrm{TP}+\mathrm{FN}}=\frac{\mathrm{TP}}{N_+}.
\]
The true negative rate (TNR), also referred to as specificity, measures instead the fraction of negative samples that are correctly classified:
\[
\mathrm{TNR}=\mathrm{Specificity}=\frac{\mathrm{TN}}{\mathrm{TN}+\mathrm{FP}}=\frac{\mathrm{TN}}{N_-}.
\]
Complementary quantities are the false positive rate (FPR) and false negative rate (FNR), which quantify the fraction of incorrect predictions within each class:
\[
\mathrm{FPR}=\frac{\mathrm{FP}}{\mathrm{FP}+\mathrm{TN}}=1-\mathrm{TNR},\qquad
\mathrm{FNR}=\frac{\mathrm{FN}}{\mathrm{FN}+\mathrm{TP}}=1-\mathrm{TPR}.
\]

A widely used global indicator is the overall accuracy, defined as the fraction of correctly classified samples over the entire dataset, i.e.
\[
\mathrm{Accuracy}=\frac{\mathrm{TP}+\mathrm{TN}}{N}.
\]
However, accuracy can be misleading in the presence of class imbalance, since a classifier may obtain a large value simply by favoring the most populated class. For this reason we also report the balanced accuracy, defined as the arithmetic mean of the per-class recalls, i.e.
\[
\mathrm{BA}=\frac{1}{2}\left(\frac{\mathrm{TP}}{\mathrm{TP}+\mathrm{FN}}+\frac{\mathrm{TN}}{\mathrm{TN}+\mathrm{FP}}\right)
=\frac{1}{2}\left(\mathrm{TPR}+\mathrm{TNR}\right).
\]
This quantity weighs the two classes symmetrically and therefore provides a more robust measure of performance when the dataset is unbalanced.

Another important indicator is precision, also called positive predictive value (PPV), which quantifies the reliability of positive predictions:
\[
\mathrm{Precision}=\mathrm{PPV}=\frac{\mathrm{TP}}{\mathrm{TP}+\mathrm{FP}}.
\]
The analogous quantity for the negative class is the negative predictive value (NPV), defined as
\[
\mathrm{NPV}=\frac{\mathrm{TN}}{\mathrm{TN}+\mathrm{FN}}.
\]
While recall measures the ability of the classifier to detect positive samples, precision quantifies how trustworthy the predicted positives are, and the two quantities are therefore often considered together when evaluating classification performance.

All the metrics introduced above are defined with respect to a given decision rule mapping the continuous model output $y$ to a discrete prediction $\hat t$. Changing the threshold (or, more generally, the decision rule) modifies the values of $\mathrm{TP}$, $\mathrm{TN}$, $\mathrm{FP}$, and $\mathrm{FN}$ and therefore affects all derived quantities. In the present work the sign of the deliberative index is used as the decision criterion for clarity and simplicity, although alternative thresholds can be introduced to tune the trade-off between precision and recall depending on the application and on the relative cost associated with false positives and false negatives.

\section{Performance Scaling of EQCM and Comparison with Extreme Learning Machines}
\label{AppC}

To further characterize the performance of the proposed
EQCM architecture, we investigate its dependence on the
dimension of the quantum feature vector and compare the
results with those obtained using an optimized classical
Extreme Learning Machine (ELM). This comparison is
particularly meaningful because both architectures
belong to the extreme learning paradigm, in which a
fixed nonlinear feature map is combined with a
trainable linear readout.

The classical ELM considered here consists of a
single hidden layer with $M$ neurons, hyperbolic
tangent activation functions, and randomly
initialized input weights and biases that remain
fixed during training. For a classical input vector
$\mathbf{z}$, the activation of the $j$-th hidden
neuron is given by
\begin{equation}
h_j(\mathbf{z}) =
\tanh\left(\mathbf{a}_j\cdot\mathbf{z}+b_j\right),
\end{equation}
where $\mathbf{a}_j$ and $b_j$ denote the fixed
random parameters. The continuous output is
obtained through a linear readout,
\begin{equation}
y_{\mathrm{ELM}}(\mathbf{z}) =
\sum_{j=1}^{M}\beta_j h_j(\mathbf{z}),
\end{equation}
whose coefficients $\beta_j$ are determined by ridge
regression. The classical benchmark is obtained
by a standard grid-search optimization of the ELM hyperparameters, while
keeping the underlying extreme learning
architecture unchanged. Increasing $M$ therefore
enlarges the space of nonlinear features available
to the linear readout.

For the quantum architecture, we consider the
hardware-compatible implementation introduced
in Sec.~\ref{sec:HF}, consisting of a seven-qubit Ising chain
with local interactions. The dynamical attention mechanism
is switched off, $g_1=g_2=0$, so that the
evolution is governed by the fixed,
input-independent Hamiltonian $H_0$.
The quantum feature vector is constructed by
progressively retaining an increasing number
of observables from the set introduced in
Sec.~\ref{sec:HF}, preserving their original ordering.
This procedure changes the number of features
available to the linear readout without
modifying the input encoding or the underlying
quantum dynamics.

\begin{figure}[htbp]
    \centering
    \includegraphics[width=0.80\linewidth]{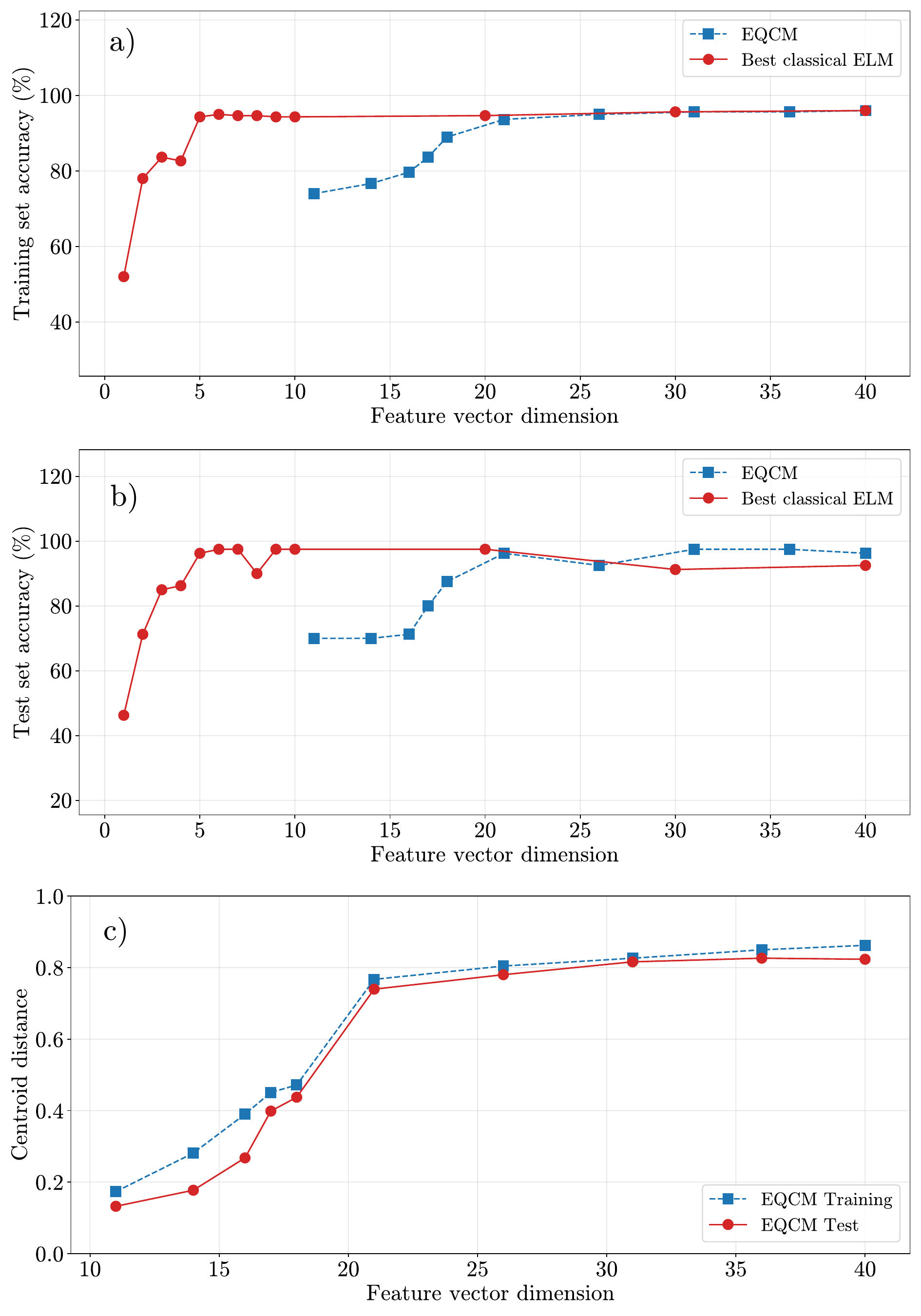}
    \caption{Scaling of the EQCM performance with the feature vector dimension in comparison with a classical Extreme Learning Machine (ELM). 
(a) Training-set accuracy for the hardware-compatible EQCM implementation without dynamical attention (blue squares) and a classical ELM with random input weights, hyperbolic tangent activation functions, and ridge-regression readout (red circles). The horizontal axis represents the number of retained quantum features for EQCM and the number of hidden neurons for ELM.
(b) Corresponding test-set accuracy. The classical ELM achieves high accuracy with fewer parameters, while EQCM reaches comparable performance as the feature dimension increases.
(c) Distance between the class-conditional centroids of the EQCM deliberative-index distributions for training and test sets. The centroid distance increases with the feature dimension, approaching the nominal target separation $t_+-t_-=1$.}

    \label{fig5}
\end{figure}

Both architectures are evaluated on the same
classification task, using identical training
and test datasets and the same dichotomic
input representation. Moreover, in both cases
learning is restricted to ridge regression
acting on fixed nonlinear features. The
comparison therefore provides a controlled
assessment of the classification performance
as a function of the readout dimension,
although it does not constitute a comparison
of the total computational or hardware
resources required by the two architectures.

Figure~\ref{fig5}(a) reports the training-set
accuracy as a function of the number of
retained quantum features for EQCM and hidden
neurons for the classical ELM. The optimized
ELM reaches its highest accuracy with a
relatively small number of neurons, whereas
the EQCM exhibits a more gradual improvement
as additional observables are included.
Nevertheless, the quantum architecture
eventually reaches an accuracy comparable
to that of the optimized classical benchmark.
A similar behaviour is observed on the test
set, as shown in Fig.~\ref{fig5}(b),
demonstrating that the improvement associated
with the larger quantum feature space is
preserved on unseen data.

Further insight into the learning process is
obtained by examining the separation between
the class-conditional distributions of the
continuous deliberative index. Denoting by
$\bar{y}_{+}$ and $\bar{y}_{-}$ the mean values
of the deliberative index evaluated on
samples belonging to the two classes,
respectively, we define the centroid
distance as
\begin{equation}
d=|\bar{y}_{+}-\bar{y}_{-}|.
\end{equation}
For the nominal targets $t_{\pm}=\pm 0.5$,
the corresponding separation is equal to one.
As shown in Fig.~\ref{fig5}(c), the centroid
distance increases with the number of
quantum features for both training and test
datasets, reflecting the progressive
development of class discrimination.
The actual centroid separation need not
coincide with the nominal target difference,
since ridge regression minimizes the
regularized prediction error rather than
enforcing prescribed class averages.
Moreover, the resulting classification
accuracy depends not only on the centroid
separation but also on the dispersion and
overlap of the corresponding distributions.

It is important to emphasize that the
scaling investigated here concerns the
dimension of the feature representation
at fixed input size. In the present EQCM
implementation, each qubit encodes one
element of the dichotomic input vector.
Consequently, increasing the number of
qubits would also require increasing the
length of the input strings, thereby
modifying the classification task and
the statistical correlations to be learned.
Such a procedure would not provide a
controlled assessment of performance
scaling under otherwise unchanged
conditions. By contrast, varying the
number of retained quantum observables
allows us to investigate the effect of
the readout dimension while preserving
the same physical dynamics and
classification problem.

The results also reveal an interesting
difference between the two architectures.
The optimized classical ELM achieves
its highest accuracy with fewer features,
suggesting a more economical
representation in terms of readout
dimension for the present benchmark.
However, the features entering the two
models have a substantially different
structure. In the classical ELM, hidden
neurons generate nonlinear transformations
belonging to the same functional family,
with diversity introduced through their
random parameters. In contrast, the
EQCM features correspond to physically
distinct quantum observables, including
local expectation values and two-qubit
correlations, which probe different
aspects of the evolved quantum state.
Their contributions to classification
are therefore not equivalent, and the
progressive inclusion of observables
in a prescribed order need not provide
the most efficient feature selection.
The observed difference in scaling
should consequently be interpreted
in relation to the specific feature
representations adopted here, rather
than as evidence of a general
superiority of one architecture.

Beyond classification accuracy,
the EQCM approach offers distinctive
conceptual and operational properties.
In particular, the learned readout
defines an explicit Hermitian
deliberative operator, whose
decomposition into physical
observables provides a direct route
to interpreting the learned
decision criterion. This operator
has a natural meaning within the
quantum cognition framework,
where internal categories are
represented by observables and
deliberative outcomes emerge as
expectation values. Furthermore,
the hardware-compatible EQCM
can be implemented using local
quantum interactions and
experimentally accessible
measurements, providing a
direct connection between the
learning architecture and its
possible physical realization.

Overall, the present benchmark
demonstrates that the EQCM can
attain classification accuracy
comparable to that of a
hyperparameter-optimized classical
ELM, while retaining its
operator-based interpretability,
its connection to quantum
cognition, and its compatibility
with quantum hardware. These
features constitute complementary
motivations for the proposed
architecture, independently of
any claim of quantum computational
advantage.

\ack{The research activity of F R received support from the PNRR MUR Project PE0000023—NQSTI,
through the cascade-funded Projects SPUNTO (CUP E63C22002180006, D.D. MUR No. 1564/2022) and
TOPQIN (CUP J13C22000680006, D.D. MUR No. 1243/2022).
J S acknowledges the contribution from PRIN (Progetti di Rilevante Interesse Nazionale) TURBIMECS, Grant No. 2022S3RSCT CUP H53D23001630006, and PNRR MUR Project PE0000023—NQSTI through the secondary Projects ‘ThAnQ’ J13C22000680006.
Open access funding provided by CRUI-CARE Agreement.}


\roles{Francesco Romeo: Conceptualization (lead), Data curation (lead), Formal analysis (lead), Funding acquisition (lead),
Investigation (lead), Methodology (lead), Supervision (lead), Writing – original draft (lead)\\
\noindent Jacopo Settino: Conceptualization (lead), Investigation (lead), Validation (lead), Visualization (lead), Writing – review \& editing (lead)}

\data{All data are available in the submitted manuscript. The data that support the findings of this study are available upon reasonable request from the authors.}


\bibliographystyle{unsrt} 
\bibliography{references}

@article{rosenblatt1958perceptron,
  title={The perceptron: a probabilistic model for information storage and organization in the brain.},
  author={Rosenblatt, Frank},
  journal={Psychological review},
  volume={65},
  number={6},
  pages={386},
  year={1958},
  publisher={American Psychological Association}
}

@article{hopfield1982neural,
  title={Neural networks and physical systems with emergent collective computational abilities.},
  author={Hopfield, John J},
  journal={Proceedings of the national academy of sciences},
  volume={79},
  number={8},
  pages={2554--2558},
  year={1982}
}

@article{amit1985spin,
  title={Spin-glass models of neural networks},
  author={Amit, Daniel J and Gutfreund, Hanoch and Sompolinsky, Haim},
  journal={Physical Review A},
  volume={32},
  number={2},
  pages={1007},
  year={1985},
  publisher={APS}
}

@article{ackley1985learning,
  title={A learning algorithm for Boltzmann machines},
  author={Ackley, David H and Hinton, Geoffrey E and Sejnowski, Terrence J},
  journal={Cognitive science},
  volume={9},
  number={1},
  pages={147--169},
  year={1985},
  publisher={Elsevier}
}

@article{mcculloch1943logical,
  title={A logical calculus of the ideas immanent in nervous activity},
  author={McCulloch, Warren S and Pitts, Walter},
  journal={The bulletin of mathematical biophysics},
  volume={5},
  number={4},
  pages={115--133},
  year={1943},
  publisher={Springer}
}

@article{caianiello1961outline,
  title={Outline of a theory of thought-processes and thinking machines},
  author={Caianiello, Eduardo R},
  journal={Journal of theoretical biology},
  volume={1},
  number={2},
  pages={204--235},
  year={1961},
  publisher={Elsevier}
}

@misc{NobelPhysics2024Background,
  author = {{Royal Swedish Academy of Sciences}},
  title  = {Scientific Background to the Nobel Prize in Physics 2024: For foundational discoveries and inventions that enable machine learning with artificial neural networks},
  year   = {2024},
  note   = {Nobel Prize Scientific Background},
  url    = {https://www.nobelprize.org/uploads/2024/09/advanced-physicsprize2024.pdf}
}

@article{little1975statistical,
  title={A statistical theory of short and long term memory},
  author={Little, WA and Shaw, Gordon L},
  journal={Behavioral biology},
  volume={14},
  number={2},
  pages={115--133},
  year={1975},
  publisher={Elsevier}
}

@article{willshaw1969non,
  title={Non-holographic associative memory},
  author={Willshaw, David J and Buneman, O Peter and Longuet-Higgins, Hugh Christopher},
  journal={Nature},
  volume={222},
  number={5197},
  pages={960--962},
  year={1969},
  publisher={Nature Publishing Group UK London}
}

@article{kohonen2009correlation,
  title={Correlation matrix memories},
  author={Kohonen, Teuvo},
  journal={IEEE transactions on computers},
  volume={100},
  number={4},
  pages={353--359},
  year={2009},
  publisher={IEEE}
}

@article{hopfield1984neurons,
  title={Neurons with graded response have collective computational properties like those of two-state neurons.},
  author={Hopfield, John J},
  journal={Proceedings of the national academy of sciences},
  volume={81},
  number={10},
  pages={3088--3092},
  year={1984}
}

@article{amit1985storing,
  title={Storing infinite numbers of patterns in a spin-glass model of neural networks},
  author={Amit, Daniel J and Gutfreund, Hanoch and Sompolinsky, Haim},
  journal={Physical review letters},
  volume={55},
  number={14},
  pages={1530},
  year={1985},
  publisher={APS}
}

@article{lecun2015deep,
  title={Deep learning},
  author={LeCun, Yann and Bengio, Yoshua and Hinton, Geoffrey},
  journal={nature},
  volume={521},
  number={7553},
  pages={436--444},
  year={2015},
  publisher={Nature Publishing Group UK London}
}

@article{schmidhuber2015deep,
  title={Deep learning in neural networks: An overview},
  author={Schmidhuber, J{\"u}rgen},
  journal={Neural networks},
  volume={61},
  pages={85--117},
  year={2015},
  publisher={Elsevier}
}

@article{rumelhart1986learning,
  title={Learning representations by back-propagating errors},
  author={Rumelhart, David E and Hinton, Geoffrey E and Williams, Ronald J},
  journal={nature},
  volume={323},
  number={6088},
  pages={533--536},
  year={1986},
  publisher={Nature Publishing Group UK London}
}

@article{schwartz2020green,
  title={Green ai},
  author={Schwartz, Roy and Dodge, Jesse and Smith, Noah A and Etzioni, Oren},
  journal={Communications of the ACM},
  volume={63},
  number={12},
  pages={54--63},
  year={2020},
  publisher={ACM New York, NY, USA}
}

@article{henderson2020towards,
  title={Towards the systematic reporting of the energy and carbon footprints of machine learning},
  author={Henderson, Peter and Hu, Jieru and Romoff, Joshua and Brunskill, Emma and Jurafsky, Dan and Pineau, Joelle},
  journal={Journal of machine learning research},
  volume={21},
  number={248},
  pages={1--43},
  year={2020}
}

@article{rudin2019stop,
  title={Stop explaining black box machine learning models for high stakes decisions and use interpretable models instead},
  author={Rudin, Cynthia},
  journal={Nature machine intelligence},
  volume={1},
  number={5},
  pages={206--215},
  year={2019},
  publisher={Nature Publishing Group UK London}
}

@article{maass2002real,
  title={Real-time computing without stable states: A new framework for neural computation based on perturbations},
  author={Maass, Wolfgang and Natschl{\"a}ger, Thomas and Markram, Henry},
  journal={Neural computation},
  volume={14},
  number={11},
  pages={2531--2560},
  year={2002},
  publisher={MIT Press}
}

@article{jaeger2004harnessing,
  title={Harnessing nonlinearity: Predicting chaotic systems and saving energy in wireless communication},
  author={Jaeger, Herbert and Haas, Harald},
  journal={science},
  volume={304},
  number={5667},
  pages={78--80},
  year={2004},
  publisher={American Association for the Advancement of Science}
}

@article{huang2006extreme,
  title={Extreme learning machine: theory and applications},
  author={Huang, Guang-Bin and Zhu, Qin-Yu and Siew, Chee-Kheong},
  journal={Neurocomputing},
  volume={70},
  number={1-3},
  pages={489--501},
  year={2006},
  publisher={Elsevier}
}

@article{tanaka2019recent,
  title={Recent advances in physical reservoir computing: A review},
  author={Tanaka, Gouhei and Yamane, Toshiyuki and H{\'e}roux, Jean Benoit and Nakane, Ryosho and Kanazawa, Naoki and Takeda, Seiji and Numata, Hidetoshi and Nakano, Daiju and Hirose, Akira},
  journal={Neural Networks},
  volume={115},
  pages={100--123},
  year={2019},
  publisher={Elsevier}
}

@article{Fujii2017,
   author = {Keisuke Fujii and Kohei Nakajima},
   doi = {https://doi.org/10.1103/PhysRevApplied.8.024030},
   issn = {23317019},
   issue = {2},
   journal = {Physical Review Applied},
   month = {8},
   pages = {024030},
   publisher = {American Physical Society},
   title = {Harnessing disordered-ensemble quantum dynamics for machine learning},
   volume = {8},
   url = {https://journals.aps.org/prapplied/abstract/10.1103/PhysRevApplied.8.024030},
   year = {2017}
}

@article{Schuld2021,
   author = {Maria Schuld and Ryan Sweke and Johannes Jakob Meyer},
   doi = {10.1103/PHYSREVA.103.032430/SUPPLEMENTAL_MATERIAL.PDF},
   issn = {24699934},
   issue = {3},
   journal = {Physical Review A},
   month = {3},
   pages = {032430},
   publisher = {American Physical Society},
   title = {Effect of data encoding on the expressive power of variational quantum-machine-learning models},
   volume = {103},
   url = {https://journals.aps.org/pra/abstract/10.1103/PhysRevA.103.032430},
   year = {2021}
}

@article{Xiong2025,
   author = {Weijie Xiong and Giorgio Facelli and Mehrad Sahebi and Owen Agnel and Thiparat Chotibut and Supanut Thanasilp and Zoë Holmes},
   doi = {10.1007/S42484-025-00239-7/FIGURES/9},
   issn = {25244914},
   issue = {1},
   journal = {Quantum Machine Intelligence},
   month = {6},
   pages = {1-33},
   publisher = {Springer Science and Business Media Deutschland GmbH},
   title = {On fundamental aspects of quantum extreme learning machines},
   volume = {7},
   url = {https://link.springer.com/article/10.1007/s42484-025-00239-7},
   year = {2025}
}

@article{Settino2024,
  title = {Memory-augmented hybrid quantum reservoir computing},
  author = {Settino, J. and Salatino, L. and Mariani, L. and D'Amore, F. and Channab, M. and Bozzolo, L. and Vallisa, S. and Barill\`a, P. and Policicchio, A. and Lo Gullo, N. and Giordano, A. and Mastroianni, C. and Plastina, F.},
  journal = {Phys. Rev. Appl.},
  volume = {24},
  issue = {2},
  pages = {024019},
  numpages = {11},
  year = {2025},
  month = {Aug},
  publisher = {American Physical Society},
  doi = {10.1103/wzwv-7rk2},
  url = {https://link.aps.org/doi/10.1103/wzwv-7rk2}
}

@article{Salatino2026,
  title={Forecasting low-dimensional turbulence via multi-dimensional hybrid quantum reservoir computing},
  author={Salatino, L and Mariani, L and Giordano, A and D’Amore, F and Mastroianni, C and Pontieri, L and Vinci, A and Gencarelli, C and Primavera, L and Plastina, F and others},
  journal={Chaos: An Interdisciplinary Journal of Nonlinear Science},
  volume={36},
  number={8},
  year={2026},
  publisher={AIP Publishing}
}

@article{Ghosh2021,
   author = {Sanjib Ghosh and Kohei Nakajima and Tanjung Krisnanda and Keisuke Fujii and Timothy C.H. Liew},
   doi = {10.1002/QUTE.202100053},
   issn = {2511-9044},
   issue = {9},
   journal = {Advanced Quantum Technologies},
   month = {9},
   pages = {2100053},
   publisher = {John Wiley \& Sons, Ltd},
   title = {Quantum Neuromorphic Computing with Reservoir Computing Networks},
   volume = {4},
   url = {https://onlinelibrary.wiley.com/doi/full/10.1002/qute.202100053 https://onlinelibrary.wiley.com/doi/abs/10.1002/qute.202100053 https://onlinelibrary.wiley.com/doi/10.1002/qute.202100053},
   year = {2021}
}

@article{Mujal2021,
   author = {Pere Mujal and Rodrigo Martínez-Peña and Johannes Nokkala and Jorge García-Beni and Gian Luca Giorgi and Miguel C. Soriano and Roberta Zambrini},
   doi = {10.1002/QUTE.202100027},
   issn = {2511-9044},
   issue = {8},
   journal = {Advanced Quantum Technologies},
   month = {8},
   pages = {2100027},
   publisher = {John Wiley \& Sons, Ltd},
   title = {Opportunities in Quantum Reservoir Computing and Extreme Learning Machines},
   volume = {4},
   url = {https://onlinelibrary.wiley.com/doi/full/10.1002/qute.202100027 https://onlinelibrary.wiley.com/doi/abs/10.1002/qute.202100027 https://onlinelibrary.wiley.com/doi/10.1002/qute.202100027},
   year = {2021}
}

@article{Finocchio2024,
   author = {Giovanni Finocchio and Jean Anne C. Incorvia and Joseph S. Friedman and Qu Yang and Anna Giordano and Julie Grollier and Hyunsoo Yang and Florin Ciubotaru and Andrii V. Chumak and Azad J. Naeemi and Sorin D. Cotofana and Riccardo Tomasello and Christos Panagopoulos and Mario Carpentieri and Peng Lin and Gang Pan and J. Joshua Yang and Aida Todri-Sanial and Gabriele Boschetto and Kremena Makasheva and Vinod K. Sangwan and Amit Ranjan Trivedi and Mark C. Hersam and Kerem Y. Camsari and Peter L. McMahon and Supriyo Datta and Belita Koiller and Gabriel H. Aguilar and Guilherme P. Temporão and Davi R. Rodrigues and Satoshi Sunada and Karin Everschor-Sitte and Kosuke Tatsumura and Hayato Goto and Vito Puliafito and Johan Åkerman and Hiroki Takesue and Massimiliano Di Ventra and Yuriy V. Pershin and Saibal Mukhopadhyay and Kaushik Roy and I. Ting Wang and Wang Kang and Yao Zhu and Brajesh Kumar Kaushik and Jennifer Hasler and Samiran Ganguly and Avik W. Ghosh and William Levy and Vwani Roychowdhury and Supriyo Bandyopadhyay},
   doi = {10.1088/2399-1984/AD299A},
   issn = {2399-1984},
   issue = {1},
   journal = {Nano Futures},
   month = {3},
   pages = {012001},
   publisher = {IOP Publishing},
   title = {Roadmap for unconventional computing with nanotechnology},
   volume = {8},
   url = {https://iopscience.iop.org/article/10.1088/2399-1984/ad299a https://iopscience.iop.org/article/10.1088/2399-1984/ad299a/meta},
   year = {2024}
}

@article{Gyurik2025,
  title={From quantum feature maps to quantum reservoir computing: an applicative perspective},
  author={Gyurik, Casper and Wudarski, Filip and Philip, Evan John and Sannia, Antonio and Sadeghi, Hossein and Kyriienko, Oleksandr and Venturelli, Davide and Gentile, Antonio Andrea},
  journal={Philosophical Transactions of the Royal Society A: Mathematical, Physical and Engineering Sciences},
  volume={384},
  number={2315},
  year={2026},
  publisher={The Royal Society}
}

@article{Ghosh2021a,
   author = {Tanjung
and Paterek Tomasz
and Liew Timothy C H Ghosh Sanjib
and Krisnanda},
   doi = {10.1038/s42005-021-00606-3},
   issn = {2399-3650},
   issue = {1},
   journal = {Communications Physics},
   month = {5},
   pages = {105},
   title = {Realising and compressing quantum circuits with quantum reservoir computing},
   volume = {4},
   url = {https://doi.org/10.1038/s42005-021-00606-3},
   year = {2021}
}

@article{Settino2026,
  title={Topology-enhanced superconducting qubit networks for in-sensor quantum information processing},
  author={Settino, J and Luciano, GG and Di Bartolomeo, A and Silvestrini, P and Lisitskiy, M and Ruggiero, B and Romeo, F},
  journal={Quantum Science and Technology},
  volume={11},
  number={1},
  pages={015019},
  year={2026},
  publisher={IOP Publishing}
}

@article{Romeo2026,
  title={Probing graph topology from local quantum measurements},
  author={Romeo, Francesco and Settino, Jacopo},
  journal={Quantum Science and Technology},
  volume={11},
  number={1},
  pages={01LT01},
  year={2026},
  publisher={IOP Publishing}
}

@book{BusemeyerBruza2012,
  author    = {Jerome R. Busemeyer and Peter D. Bruza},
  title     = {Quantum Models of Cognition and Decision},
  publisher = {Cambridge University Press},
  address   = {Cambridge},
  year      = {2012},
  isbn      = {9780521895691}
}

@article{Pothos2013,
  author  = {Pothos, E M and Busemeyer, J R},
  title   = {Can quantum probability provide a new direction for cognitive modeling?},
  journal = {Behav. Brain Sci.},
  year    = {2013},
  volume  = {36},
  pages   = {255--274}
}

@article{Khrennikov2004,
  author  = {Khrennikov, Andrei},
  title   = {Information dynamics in cognitive, psychological, social, and anomalous phenomena},
  journal = {Fundamental Theories of Physics},
  year    = {2004},
  volume  = {138}
}

@article{Aerts2009,
  author  = {Aerts, Diederik},
  title   = {Quantum structure in cognition},
  journal = {J. Math. Psychol.},
  year    = {2009},
  volume  = {53},
  pages   = {314--348}
}

@article{Accardi1981,
  author  = {Accardi, Luigi},
  title   = {Topics in quantum probability},
  journal = {Phys. Rep.},
  year    = {1981},
  volume  = {77},
  pages   = {169--192}
}

@article{Gudder1993,
  author  = {Gudder, Stanley},
  title   = {Foundations of quantum probability},
  journal = {Int. J. Theor. Phys.},
  year    = {1993},
  volume  = {32},
  pages   = {1747--1762}
}

@article{Geller2022,
  author  = {Geller, Michael R. and Arrasmith, Andrew and Holmes, Zoe and Yan, Bin and Coles, Patrick J. and Sornborger, Andrew},
  title   = {Quantum simulation of operator spreading in the chaotic Ising model},
  journal = {Phys. Rev. E},
  year    = {2022},
  volume  = {105},
  pages   = {035302}
}

@article{DeLorenzis2025,
  author  = {De Lorenzis, A. and Casado, M. P. and Estarellas, M. P. and Lo Gullo, N. and Lux, T. and Plastina, F. and Riera, A. and Settino, J.},
  title   = {Harnessing quantum extreme learning machines for image classification},
  journal = {Phys. Rev. Applied},
  year    = {2025},
  volume  = {23},
  pages   = {044024}
}

@article{Jaynes57,
  title = {Information Theory and Statistical Mechanics},
  author = {Jaynes, E. T.},
  journal = {Phys. Rev.},
  volume = {106},
  issue = {4},
  pages = {620--630},
  numpages = {0},
  year = {1957},
  month = {May},
  publisher = {American Physical Society},
  doi = {10.1103/PhysRev.106.620},
  url = {https://link.aps.org/doi/10.1103/PhysRev.106.620}
}

@book{haake1991quantum,
    author = "Haake, Fritz",
    title = "{Quantum Signatures of Chaos}",
    doi = "10.1007/978-3-642-05428-0",
    isbn = "978-3-642-26330-9, 978-3-642-05428-0",
    publisher = "Springer",
    address = "Berlin",
    series = "Springer Series in Synergetics",
    year = "2010"
}

@article{beggs2003neuronal,
  title={Neuronal avalanches in neocortical circuits},
  author={Beggs, John M and Plenz, Dietmar},
  journal={Journal of neuroscience},
  volume={23},
  number={35},
  pages={11167--11177},
  year={2003},
  publisher={Society for Neuroscience}
}

@article{chialvo2010emergent,
  title={Emergent complex neural dynamics},
  author={Chialvo, Dante R},
  journal={Nature physics},
  volume={6},
  number={10},
  pages={744--750},
  year={2010},
  publisher={Nature Publishing Group UK London}
}

@article{shew2013functional,
  title={The functional benefits of criticality in the cortex},
  author={Shew, Woodrow L and Plenz, Dietmar},
  journal={The neuroscientist},
  volume={19},
  number={1},
  pages={88--100},
  year={2013},
  publisher={SAGE Publications Sage CA: Los Angeles, CA}
}

@article{vaswani2017attention,
  title={Attention is all you need},
  author={Vaswani, Ashish and Shazeer, Noam and Parmar, Niki and Uszkoreit, Jakob and Jones, Llion and Gomez, Aidan N and Kaiser, {\L}ukasz and Polosukhin, Illia},
  journal={Advances in neural information processing systems},
  volume={30},
  year={2017}
}

@article{huang2020predicting,
  title={Predicting many properties of a quantum system from very few measurements},
  author={Huang, Hsin-Yuan and Kueng, Richard and Preskill, John},
  journal={Nature Physics},
  volume={16},
  number={10},
  pages={1050--1057},
  year={2020},
  publisher={Nature Publishing Group UK London}
}

@article{havlivcek2019supervised,
  title={Supervised learning with quantum-enhanced feature spaces},
  author={Havl{\'\i}{\v{c}}ek, Vojt{\v{e}}ch and C{\'o}rcoles, Antonio D and Temme, Kristan and Harrow, Aram W and Kandala, Abhinav and Chow, Jerry M and Gambetta, Jay M},
  journal={Nature},
  volume={567},
  number={7747},
  pages={209--212},
  year={2019},
  publisher={Nature Publishing Group UK London}
}

@article{hoerl1970ridge,
  title={Ridge regression: Biased estimation for nonorthogonal problems},
  author={Hoerl, Arthur E and Kennard, Robert W},
  journal={Technometrics},
  volume={12},
  number={1},
  pages={55--67},
  year={1970},
  publisher={Taylor \& Francis}
}

@article{shannon1948mathematical,
  title={A mathematical theory of communication},
  author={Shannon, Claude Elwood},
  journal={The Bell system technical journal},
  volume={27},
  number={3},
  pages={379--423},
  year={1948},
  publisher={Nokia Bell Labs}
}

@book{pearson1904theory,
  title={On the theory of contingency and its relation to association and normal correlation},
  author={Pearson, Karl},
  volume={1},
  year={1904},
  publisher={Cambridge University Press}
}

@article{miller1955analysis,
  title={An analysis of perceptual confusions among some English consonants},
  author={Miller, George A and Nicely, Patricia E},
  journal={The Journal of the Acoustical Society of America},
  volume={27},
  number={2},
  pages={338--352},
  year={1955},
  publisher={Acoustical Society of America}
}

@Book{zipf1949human,
author={Zipf, George Kingsley},
title={Human behavior and the principle of least effort.},
year={1949},
publisher={Addison-Wesley Press},
address={Oxford,  England},
pages={xi, 573-xi, 573}
}

@article{nespor2003different,
  title={On the different roles of vowels and consonants in speech processing and language acquisition},
  author={Nespor, Marina and Pe{\~n}a, Marcela and Mehler, Jacques},
  journal={Lingue e linguaggio},
  volume={2},
  number={2},
  pages={203--230},
  year={2003},
  publisher={Societ{\`a} editrice il Mulino}
}

@article{mazzarisi2021maximal,
  title={Maximal diversity and Zipf’s law},
  author={Mazzarisi, Onofrio and de Azevedo-Lopes, Amanda and Arenzon, Jeferson J and Corberi, Federico},
  journal={Physical review letters},
  volume={127},
  number={12},
  pages={128301},
  year={2021},
  publisher={APS}
}

@book{KlirYuan1995,
  author = {Klir, George J. and Yuan, Bo},
  title = {Fuzzy Sets and Fuzzy Logic: Theory and Applications},
  publisher = {Prentice Hall},
  year = {1995}
}

@article{PhysRevLett.60.1351,
  title = {How the result of a measurement of a component of the spin of a spin-1/2 particle can turn out to be 100},
  author = {Aharonov, Yakir and Albert, David Z. and Vaidman, Lev},
  journal = {Phys. Rev. Lett.},
  volume = {60},
  issue = {14},
  pages = {1351--1354},
  numpages = {0},
  year = {1988},
  month = {Apr},
  publisher = {American Physical Society},
  doi = {10.1103/PhysRevLett.60.1351},
  url = {https://link.aps.org/doi/10.1103/PhysRevLett.60.1351}
}

@article{RevModPhys.86.307,
  title = {Colloquium: Understanding quantum weak values: Basics and applications},
  author = {Dressel, Justin and Malik, Mehul and Miatto, Filippo M. and Jordan, Andrew N. and Boyd, Robert W.},
  journal = {Rev. Mod. Phys.},
  volume = {86},
  issue = {1},
  pages = {307--316},
  numpages = {10},
  year = {2014},
  month = {Mar},
  publisher = {American Physical Society},
  doi = {10.1103/RevModPhys.86.307},
  url = {https://link.aps.org/doi/10.1103/RevModPhys.86.307}
}

@article{Mikolov2013,
  author = {Mikolov, Tomas and Sutskever, Ilya and Chen, Kai and Corrado, Greg and Dean, Jeffrey},
  title = {Distributed representations of words and phrases and their compositionality},
  journal = {Advances in Neural Information Processing Systems},
  volume = {26},
  year = {2013}
}

@article{Karniadakis2021,
  author = {Karniadakis, George E. and Kevrekidis, Ioannis G. and Lu, Lu and Perdikaris, Paris and Wang, Sifan and Yang, Liu},
  title = {Physics-informed machine learning},
  journal = {Nature Reviews Physics},
  volume = {3},
  number = {6},
  pages = {422--440},
  year = {2021},
  doi = {10.1038/s42254-021-00314-5},
  url = {https://doi.org/10.1038/s42254-021-00314-5}
}

@book{Tao2012,
  author = {Tao, Terence},
  title = {Topics in Random Matrix Theory},
  series = {Graduate Studies in Mathematics},
  volume = {132},
  publisher = {American Mathematical Society},
  address = {Providence, RI},
  year = {2012},
  isbn = {978-0-8218-7430-1}
}

@article{Sannia2024,
  author = {Sannia, Antonio and Mart{\'\i}nez-Pe{\~n}a, Rodrigo and Soriano, Miguel C. and Giorgi, Gian Luca and Zambrini, Roberta},
  title = {Dissipation as a resource for Quantum Reservoir Computing},
  journal = {Quantum},
  volume = {8},
  pages = {1291},
  year = {2024},
  doi = {10.22331/q-2024-03-20-1291},
  url = {https://doi.org/10.22331/q-2024-03-20-1291},
  eprint = {2212.12078},
  archivePrefix = {arXiv},
  primaryClass = {quant-ph}
}

@article{jing2025,
  title = {Quantum transport reservoir computing},
  author = {Jing, Yecheng and Wang, Pengfei and Zhang, Shuai and Zeng, Zhoujie and Liang, Shi-Jun and Chen, Wei},
  journal = {Phys. Rev. Appl.},
  volume = {24},
  issue = {4},
  pages = {044036},
  numpages = {11},
  year = {2025},
  month = {Oct},
  publisher = {American Physical Society},
  doi = {10.1103/w117-7gmd},
  url = {https://link.aps.org/doi/10.1103/w117-7gmd}
}

\end{document}